%% file: main.tex
\pdfoutput=1

\documentclass[12pt,a4paper]{article}

\usepackage{ifthen} 
\usepackage{subfigure}

\newboolean{pdflatex}
\setboolean{pdflatex}{true} 

\newboolean{articletitles}
\setboolean{articletitles}{true} 

\newboolean{uprightparticles}
\setboolean{uprightparticles}{false}

\def\paperauthors{LHCb collaboration} 
\def\paperasciititle{Measurement of the CKM angle gamma in B->D*h channels with D*->Dpi0/gamma, D->Kshh decays} 
\def\papertitle{Measurement of the CKM angle \texorpdfstring{$\gamma$}{gamma} using the \texorpdfstring{$B^\pm\to \Dstar h^\pm$ channels}{B to Dsth}}
\def\paperkeywords{{High Energy Physics}, {LHCb}} 
\def\papercopyright{\the\year\ CERN for the benefit of the LHCb collaboration} 
\def\paperlicence{CC BY 4.0 licence}
\def\paperlicenceurl{https://creativecommons.org/licenses/by/4.0/}

\input{preamble}

\usepackage[capitalise]{cleveref}

\input{extradef.tex}
\usepackage{longtable} 

\begin{document}

\renewcommand{\thefootnote}{\fnsymbol{footnote}}
\setcounter{footnote}{1}

\input{title-LHCb-PAPER}

\renewcommand{\thefootnote}{\arabic{footnote}}
\setcounter{footnote}{0}

\cleardoublepage

\pagestyle{plain} 
\setcounter{page}{1}
\pagenumbering{arabic}

\input{introduction}

\input{theory}

\input{detector}

\input{selection}

 \input{mass_fits}

 \input{cpobv}

 \input{systematic}

 \input{interpret}

\input{conclude}

\input{acknowledgements}

\input{appendix}

\addcontentsline{toc}{section}{References}

\bibliographystyle{LHCb}
\bibliography{main,standard,LHCb-PAPER,LHCb-CONF,LHCb-DP,LHCb-TDR}

\newpage
\input{Authorship_LHCb-PAPER-2023-012}

\end{document}

%% file: preamble.tex
\usepackage[top=1in, bottom=1.25in, left=1in, right=1in]{geometry}

\usepackage{microtype}
\usepackage{lineno}  
\usepackage{xspace} 
                    
\usepackage{caption}

\usepackage{graphicx}  
\usepackage{color}
\usepackage{colortbl}
\graphicspath{{./figs/}}

\usepackage{amsmath} 
\usepackage{amssymb}
\usepackage{amsfonts}
\usepackage{upgreek}

\newcommand*\patchAmsMathEnvironmentForLineno[1]{%
\expandafter\let\csname old#1\expandafter\endcsname\csname #1\endcsname
\expandafter\let\csname oldend#1\expandafter\endcsname\csname
end#1\endcsname
 \renewenvironment{#1}%
   {\linenomath\csname old#1\endcsname}%
   {\csname oldend#1\endcsname\endlinenomath}%
}
\newcommand*\patchBothAmsMathEnvironmentsForLineno[1]{%
  \patchAmsMathEnvironmentForLineno{#1}%
  \patchAmsMathEnvironmentForLineno{#1*}%
}
\AtBeginDocument{%
\patchBothAmsMathEnvironmentsForLineno{equation}%
\patchBothAmsMathEnvironmentsForLineno{align}%
\patchBothAmsMathEnvironmentsForLineno{flalign}%
\patchBothAmsMathEnvironmentsForLineno{alignat}%
\patchBothAmsMathEnvironmentsForLineno{gather}%
\patchBothAmsMathEnvironmentsForLineno{multline}%
\patchBothAmsMathEnvironmentsForLineno{eqnarray}%
}

\usepackage{hyperxmp}

\usepackage[pdftex,
            pdfauthor={\paperauthors},
            pdftitle={\paperasciititle},
            pdfkeywords={\paperkeywords},
            pdfcopyright={Copyright (C) \papercopyright},
            pdflicenseurl={\paperlicenceurl}]{hyperref}

\usepackage[colorinlistoftodos,textsize=scriptsize]{todonotes}

\usepackage[bottom,flushmargin,hang,multiple]{footmisc}

\usepackage[all]{hypcap}

\input{lhcb-symbols-def}

\usepackage{cite} 
\usepackage{mciteplus}

%% file: lhcb-symbols-def.tex
\usepackage{xspace} 
\usepackage{upgreek}

\def\lhcb   {\mbox{LHCb}\xspace}

\def\babar  {\mbox{BaBar}\xspace}
\def\belle  {\mbox{Belle}\xspace}

\def\besiii {\mbox{BESIII}\xspace}
\def\cleo   {\mbox{CLEO}\xspace}

\def\MagUp {\mbox{\em Mag\kern -0.05em Up}\xspace}

\ifthenelse{\boolean{uprightparticles}}
{

 \def\Pmu         {\ensuremath{\upmu}\xspace}

 \def\Ppi         {\ensuremath{\uppi}\xspace}

 \def\Ppsi        {\ensuremath{\uppsi}\xspace}

 \def\PDelta      {\ensuremath{\Delta}\xspace}                 
 \def\PXi         {\ensuremath{\Xi}\xspace}                 
 \def\PLambda     {\ensuremath{\Lambda}\xspace}                 
 \def\PSigma      {\ensuremath{\Sigma}\xspace}                 
 \def\POmega      {\ensuremath{\Omega}\xspace}                 
 \def\PUpsilon    {\ensuremath{\Upsilon}\xspace}
 \let\oldPi\Pi
 \def\PPi         {\ensuremath{\oldPi}\xspace}

 \def\PB      {\ensuremath{\mathrm{B}}\xspace}                 
                  
 \def\PD      {\ensuremath{\mathrm{D}}\xspace}

 \def\PK      {\ensuremath{\mathrm{K}}\xspace}

 \def\Pb      {\ensuremath{\mathrm{b}}\xspace}                 
 \def\Pc      {\ensuremath{\mathrm{c}}\xspace}                 
                  
 \def\Pe      {\ensuremath{\mathrm{e}}\xspace}

 \def\Ph      {\ensuremath{\mathrm{h}}\xspace}                 
 \def\Pi      {\ensuremath{\mathrm{i}}\xspace}

 \def\Ps      {\ensuremath{\mathrm{s}}\xspace}

 \def\thebaroffset{0.0em}
}
{

 \def\Pmu         {\ensuremath{\mu}\xspace}

 \def\Ppi         {\ensuremath{\pi}\xspace}

 \def\Ppsi        {\ensuremath{\psi}\xspace}                 
                  
 \mathchardef\PDelta="7101
 \mathchardef\PXi="7104
 \mathchardef\PLambda="7103
 \mathchardef\PSigma="7106
 \mathchardef\POmega="710A
 \mathchardef\PUpsilon="7107
 \mathchardef\PPi="7105
                  
 \def\PB      {\ensuremath{B}\xspace}                 
                  
 \def\PD      {\ensuremath{D}\xspace}

 \def\PK      {\ensuremath{K}\xspace}

 \def\Pb      {\ensuremath{b}\xspace}                 
 \def\Pc      {\ensuremath{c}\xspace}                 
                  
 \def\Pe      {\ensuremath{e}\xspace}

 \def\Ph      {\ensuremath{h}\xspace}                 
 \def\Pi      {\ensuremath{i}\xspace}

 \def\Ps      {\ensuremath{s}\xspace}

 \def\thebaroffset{0.18em}
}
\newcommand{\offsetoverline}[2][\thebaroffset]{\kern #1\overline{\kern -#1 #2}}%

\makeatletter
\ifcase \@ptsize \relax
  \newcommand{\miniscule}{\@setfontsize\miniscule{4}{5}}
\or
  \newcommand{\miniscule}{\@setfontsize\miniscule{5}{6}}
\or
  \newcommand{\miniscule}{\@setfontsize\miniscule{5}{6}}
\fi
\makeatother

\DeclareRobustCommand{\optbar}[1]{\shortstack{{\miniscule (\rule[.5ex]{1.25em}{.18mm})}
  \\ [-.7ex] $#1$}}

\def\electron   {{\ensuremath{\Pe}}\xspace}

\def\muon       {{\ensuremath{\Pmu}}\xspace}

\def\squark    {{\ensuremath{\Ps}}\xspace}

\def\cquark    {{\ensuremath{\Pc}}\xspace}

\def\bquark    {{\ensuremath{\Pb}}\xspace}

\def\hadron {{\ensuremath{\Ph}}\xspace}
\def\pion   {{\ensuremath{\Ppi}}\xspace}
\def\piz    {{\ensuremath{\pion^0}}\xspace}
\def\pip    {{\ensuremath{\pion^+}}\xspace}
\def\pim    {{\ensuremath{\pion^-}}\xspace}
\def\pipm   {{\ensuremath{\pion^\pm}}\xspace}

\def\kaon    {{\ensuremath{\PK}}\xspace}

\def\KorKbar {\kern \thebaroffset\optbar{\kern -\thebaroffset \PK}{}\xspace}

\def\Kp      {{\ensuremath{\kaon^+}}\xspace}
\def\Km      {{\ensuremath{\kaon^-}}\xspace}
\def\Kpm     {{\ensuremath{\kaon^\pm}}\xspace}

\def\KS      {{\ensuremath{\kaon^0_{\mathrm{S}}}}\xspace}

\def\Dbar    {{\ensuremath{\offsetoverline{\PD}}}\xspace}
\def\D       {{\ensuremath{\PD}}\xspace}
\def\Db      {{\ensuremath{\Dbar}}\xspace}
\def\DorDbar {\kern \thebaroffset\optbar{\kern -\thebaroffset \PD}\xspace}
\def\Dz      {{\ensuremath{\D^0}}\xspace}
\def\Dzb     {{\ensuremath{\Dbar{}^0}}\xspace}
\def\Dp      {{\ensuremath{\D^+}}\xspace}
\def\Dm      {{\ensuremath{\D^-}}\xspace}

\def\DpDm    {\ensuremath{\Dp {\kern -0.16em \Dm}}\xspace}
\def\Dstar   {{\ensuremath{\D^*}}\xspace}

\def\Dstarz  {{\ensuremath{\D^{*0}}}\xspace}
\def\Dstarzb {{\ensuremath{\Dbar{}^{*0}}}\xspace}

\def\B       {{\ensuremath{\PB}}\xspace}

\def\BorBbar {\kern \thebaroffset\optbar{\kern -\thebaroffset \PB}\xspace}

\def\Bd      {{\ensuremath{\B^0}}\xspace}

\def\BdorBdbar {\kern \thebaroffset\optbar{\kern -\thebaroffset \Bd}\xspace}
\def\Bu      {{\ensuremath{\B^+}}\xspace}
\def\Bub     {{\ensuremath{\B^-}}\xspace}
\def\Bp      {{\ensuremath{\Bu}}\xspace}
\def\Bm      {{\ensuremath{\Bub}}\xspace}
\def\Bpm     {{\ensuremath{\B^\pm}}\xspace}

\def\Bs      {{\ensuremath{\B^0_\squark}}\xspace}

\def\BsorBsbar {\kern \thebaroffset\optbar{\kern -\thebaroffset \Bs}\xspace}
\def\Bc      {{\ensuremath{\B_\cquark^+}}\xspace}

\def\psiprpr  {{\ensuremath{\Ppsi(3770)}}\xspace}

\def\Y#1S{\ensuremath{\PUpsilon{(#1S)}}\xspace}

\def\Lz          {{\ensuremath{\PLambda}}\xspace}

\def\LorLbar     {\kern \thebaroffset\optbar{\kern -\thebaroffset \PLambda}\xspace}

\def\Lb           {{\ensuremath{\Lz^0_\bquark}}\xspace}

\def\to                 {\ensuremath{\rightarrow}\xspace}

\newcommand{\mBs}{{\ensuremath{m_{\Bs}}}\xspace}

\def\CP                {{\ensuremath{C\!P}}\xspace}

\def\AT#1     {\ensuremath{A_{\mathrm{T}}^{#1}}\xspace}

\def\C#1      {\ensuremath{\mathcal{C}_{#1}}\xspace}                       
\def\Cp#1     {\ensuremath{\mathcal{C}_{#1}^{'}}\xspace}                    
\def\Ceff#1   {\ensuremath{\mathcal{C}_{#1}^{\mathrm{(eff)}}}\xspace}         
\def\Cpeff#1  {\ensuremath{\mathcal{C}_{#1}^{'\mathrm{(eff)}}}\xspace}       
\def\Ope#1    {\ensuremath{\mathcal{O}_{#1}}\xspace}                       
\def\Opep#1   {\ensuremath{\mathcal{O}_{#1}^{'}}\xspace}

\newcommand{\nospaceunit}[1]{\ensuremath{\text{#1}}}       
\newcommand{\aunit}[1]{\ensuremath{\text{\,#1}}}

\newcommand{\tev}{\aunit{Te\kern -0.1em V}\xspace}
\newcommand{\gev}{\aunit{Ge\kern -0.1em V}\xspace}
\newcommand{\mev}{\aunit{Me\kern -0.1em V}\xspace}
\newcommand{\kev}{\aunit{ke\kern -0.1em V}\xspace}
\newcommand{\ev}{\aunit{e\kern -0.1em V}\xspace}
 
\newcommand{\mevc}{\ensuremath{\aunit{Me\kern -0.1em V\!/}c}\xspace}
\newcommand{\gevc}{\ensuremath{\aunit{Ge\kern -0.1em V\!/}c}\xspace}
\newcommand{\mevcc}{\ensuremath{\aunit{Me\kern -0.1em V\!/}c^2}\xspace}
\newcommand{\gevcc}{\ensuremath{\aunit{Ge\kern -0.1em V\!/}c^2}\xspace}

\def\mum  {\ensuremath{\,\upmu\nospaceunit{m}}\xspace}

\def\fb   {\ensuremath{\aunit{fb}}\xspace}
\def\invfb   {\ensuremath{\fb^{-1}}\xspace}

\newcommand{\chisq}{\ensuremath{\chi^2}\xspace}

\newcommand{\chisqip}{\ensuremath{\chi^2_{\text{IP}}}\xspace}

\def\gsim{{~\raise.15em\hbox{$>$}\kern-.85em
          \lower.35em\hbox{$\sim$}~}\xspace}
\def\lsim{{~\raise.15em\hbox{$<$}\kern-.85em
          \lower.35em\hbox{$\sim$}~}\xspace}

\def\sPlot{\mbox{\em sPlot}\xspace}

\def\pt         {\ensuremath{p_{\mathrm{T}}}\xspace}

\def\ptot       {\ensuremath{p}\xspace}

\def\dirac      {\mbox{\textsc{Dirac}}\xspace}

\def\evtgen     {\mbox{\textsc{EvtGen}}\xspace}

\def\geant      {\mbox{\textsc{Geant4}}\xspace}

\def\photos     {\mbox{\textsc{Photos}}\xspace}

\def\pythia     {\mbox{\textsc{Pythia}}\xspace}

\def\tensorflow {\mbox{\textsc{TensorFlow}}\xspace}

\def\tell1  {TELL1\xspace}
\def\ukl1   {UKL1\xspace}

\newcommand{\ie}{\mbox{\itshape i.e.}\xspace}

\newcommand{\lhcborcid}[1]{\href{https://orcid.org/#1}{\hspace*{0.1em}\raisebox{-0.45ex}{\includegraphics[width=1em]{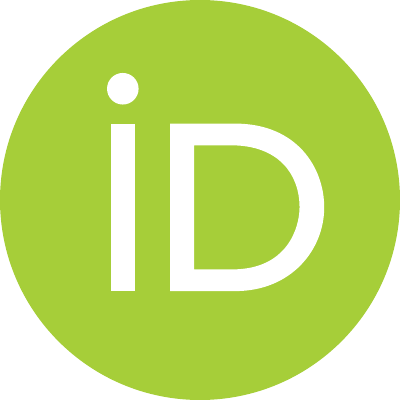}}}}

%% file: extradef.tex
\def\mdn    {{\ensuremath{m(\D \piz/\gamma)}}\xspace}
\def\mdh    {{\ensuremath{m(\D h^{\pm})}}\xspace}

\def\ckmgammaangle {{ \ensuremath{\gamma}}\xspace}

\def\ckmanglegamma {{{\rm CKM\ }\xspace {\rm angle}\   \ensuremath{\gamma}}\xspace}

\def\hp    {{\ensuremath{\hadron^+}}\xspace}
\def\hm    {{\ensuremath{\hadron^-}}\xspace}
\def\hpm    {{\ensuremath{\hadron^\pm}}\xspace}

\newcommand{\xpe}{\ensuremath{3.16}}
\newcommand{\xme}{\ensuremath{3.55}}
\newcommand{\ype}{\ensuremath{4.41}}
\newcommand{\yme}{\ensuremath{3.98}}
\newcommand{\xxie}{\ensuremath{5.00}}
\newcommand{\yxie}{\ensuremath{5.04}}

\newcommand{\xpvstat}{\ensuremath{\phantom{-}11.42\pm 3.16}}
\newcommand{\xmvstat}{\ensuremath{\phantom{}-8.91\pm 3.55}}
\newcommand{\ypvstat}{\ensuremath{\phantom{-1}3.60\pm 4.41}}
\newcommand{\ymvstat}{\ensuremath{-16.75\pm 3.98}}
\newcommand{\xxivstat}{\ensuremath{\phantom{-1}0.51\pm 5.00}}
\newcommand{\yxivstat}{\ensuremath{\phantom{-1}7.92\pm 5.04}}

\newcommand{\xpvsys}{\ensuremath{1.26}}
\newcommand{\xmvsys}{\ensuremath{2.04}}
\newcommand{\ypvsys}{\ensuremath{2.12}}
\newcommand{\ymvsys}{\ensuremath{1.48}}
\newcommand{\xxivsys}{\ensuremath{2.66}}
\newcommand{\yxivsys}{\ensuremath{3.78}}

\newcommand{\vargamma}{\ensuremath{ (69^{+13}_{-14}) ^{\circ}}}
\newcommand{\varrbdstk}{\ensuremath{0.15 \pm 0.03}}
\newcommand{\varrbdstpi}{\ensuremath{0.01 \pm 0.01}}
\newcommand{\vardeltabdstk}{\ensuremath{(311 \pm 14)^{\circ}}}
\newcommand{\vardeltabdstpi}{\ensuremath{(37 \pm 37)^{\circ}}}

%% file: title-LHCb-PAPER.tex
\begin{titlepage}
\pagenumbering{roman}

\vspace*{-1.5cm}
\centerline{\large EUROPEAN ORGANIZATION FOR NUCLEAR RESEARCH (CERN)}
\vspace*{1.5cm}
\noindent
\begin{tabular*}{\linewidth}{lc@{\extracolsep{\fill}}r@{\extracolsep{0pt}}}
\ifthenelse{\boolean{pdflatex}}
{\vspace*{-1.5cm}\mbox{\!\!\!\includegraphics[width=.14\textwidth]{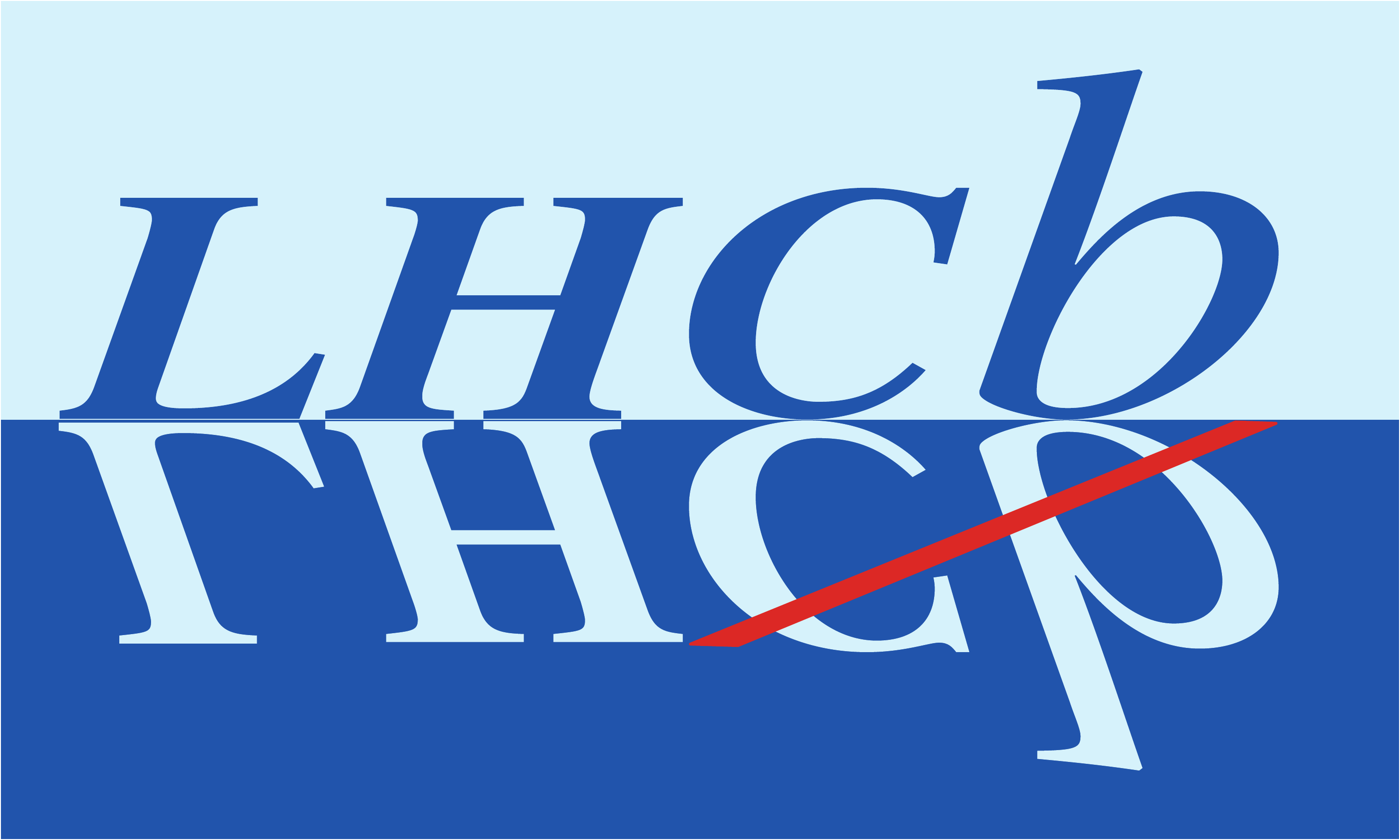}} & &}
{\vspace*{-1.2cm}\mbox{\!\!\!\includegraphics[width=.12\textwidth]{lhcb-logo.eps}} & &}
\\
 & & CERN-EP-2023-170 \\  
 & & LHCb-PAPER-2023-012 \\  
 & & \today \\ 
 & & \\

\end{tabular*}

\vspace*{3.0cm}

{\normalfont\bfseries\boldmath\huge
\begin{center}

  \papertitle 
\end{center}
}

\vspace*{1.0cm}

\begin{center}

\paperauthors\footnote{Authors are listed at the end of this paper.}
\end{center}

\begin{abstract}
  \noindent
  A measurement of the \CP-violating observables from $B^{\pm}\rightarrow \Dstar K^{\pm}$ and \mbox{$B^{\pm}\rightarrow \Dstar \pi^{\pm}$} decays is presented, where $\Dstar (D) $ is an admixture of $\Dstarz$ and $\Dstarzb$ ($D^0$ and $\bar{D}^0$) states and is reconstructed through the decay chains $ \Dstar \rightarrow D\pi^0/\gamma$ and \mbox{$D \to \KS \pi^+\pi^-/\KS K^+K^-$}. 
  The measurement is performed by analysing the signal yield variation across the $D$ decay phase space and is independent of any amplitude model. The data sample used was collected by the \lhcb experiment in proton-proton collisions and corresponds to a total integrated luminosity of 9 \invfb at centre-of-mass energies of 7, 8 and 13 \tev.
 The CKM angle $\gamma$ is determined to be $\vargamma$ using the measured \CP-violating observables. The hadronic parameters $r^{\Dstar K^{\pm}}_B, r^{\Dstar \pi^{\pm}}_B, \delta^{\Dstar K^{\pm}}_B, \delta^{\Dstar \pi^{\pm}}_B$, which are the ratios and strong phase differences between favoured and suppressed $B^{\pm}$ decays, are also reported.

\end{abstract}
\vspace*{1.0cm}
\begin{center}
  Published in JHEP12(2023)013
\end{center}

\vspace{\fill}

{\footnotesize 

\centerline{\copyright~\papercopyright. \href{\paperlicenceurl}{\paperlicence}.}}
\vspace*{2mm}

\end{titlepage}

\newpage
\setcounter{page}{2}
\mbox{~}

%% file: introduction.tex
\section{Introduction}
\label{sec:introduction}
 
The combined charge-parity (\CP) symmetry violation (\CP violation) in weak interactions in the Standard Model (SM) arises from a single, nonzero, irreducible weak phase angle in the CKM matrix elements, $V_{ij}$~\cite{PhysRevLett.10.531, 10.1143/PTP.49.652}. The CKM angle \mbox{$\gamma \equiv \textrm{arg}(-V_{ud}V^*_{ub}/V_{cd}V_{cb}^*)$}, can be described by the angles and lengths of the unitarity triangle constructed from elements of CKM matrix. This angle can be measured directly through the interference between $b\rightarrow c$ and $b\rightarrow u$ processes in tree-level decays to the same final-state particles. The theoretical uncertainty in the interpretation of these measurements in terms of $\ckmgammaangle$ is $\mathcal{O}(10^{-7})$~\cite{Brod:2013sga}. In the absence of processes beyond the SM at tree level, the measurement of $\ckmgammaangle$ is not expected to receive any other contributions. These measurements of $\ckmgammaangle$ thus provide an SM benchmark, which can be compared to ``indirect'' determinations inferred from observables that are more likely to receive contributions from new physics~\cite{Blanke:2018cya}. 
A comparison between direct and indirect determinations of $\ckmgammaangle$ serves as a test of the SM description of \CP-violation and probes possible physics beyond the SM, making over-constraining the CKM matrix an important experimental objective. The current indirectly determined value is $\ckmgammaangle= (65.7^{+1.3}_{-1.2})^{\circ}$~\cite{CKMfitter2015} and the world average of the direct measurements is $\gamma = (67\pm4)^{\circ}$~\cite{HFLAV:2022pwe}. While good agreement is seen at the current level of precision, the uncertainties in the individual direct measurements are large and can be improved upon. 

The golden modes for measuring $\gamma$ directly are the $\Bpm\to DK^{\pm}$ and $\Bpm\to D\pi^{\pm}$ family of decays, in which the $D$ meson can be reconstructed in a variety of final states offering complementary sensitivity. Though modes involving excited $D^\ast$ mesons also provide good sensitivity, thus far there have been only a few measurements with excited $D^\ast$ mesons. At \lhcb, measurements have been performed with $\Bpm\rightarrow \Dstar K^{\pm}$ and \mbox{$B^{\pm}\rightarrow \Dstar \pi^{\pm}$} (commonly denoted $B^{\pm} \rightarrow \Dstar h^{\pm}$) decays, with $\Dstar  \to D\piz/\gamma$ and the $D$ meson reconstructed using decays to two hadrons (GLW-ADS modes)~\cite{LHCb-PAPER-2020-036}. Here $\Dstar $ ($D$) denotes admixtures of $\Dstarz$ ($\Dz$) and \Dstarzb ($\Dzb$) mesons. In these analyses, the final-state neutral particles ($\piz/\gamma$) were not reconstructed and GLW-ADS modes need measurement with multiple body $D$ decay to give a single solution. The present analysis studies the same $B^{\pm}$ decays with $\Dstar\to \D\piz/\gamma, \D \to \KS \pi^+\pi^-/\KS K^+K^-$ (denoted $D\rightarrow \KS h^+h^-$) with all final-state particles being reconstructed. Furthermore, events with partially or incorrectly reconstructed $\Dstar$ particles retain some sensitivity to $\gamma$ and are included in the analysis (\cref{sec:mass_fits}). 
This measurement has also been performed by \belle and \babar with the same decay modes, employing a technique that relies on an amplitude model of the $\Dz$ decay~\cite{Poluektov:2004mf, PhysRevLett.105.121801}. Instead, a model-independent analysis, which does not rely on the amplitude model of the \D decays, is performed for this measurement.

These measurements of \CP-violation effects rely on the interference between strong and weak phases.
The rich resonance content in $\D\to\KS\hp\hm$ decays with different strong phases ensure good sensitivity to $\ckmgammaangle$ but require knowledge of these strong phases throughout the phase space.
The model-independent approach~\cite{Giri:2003ty} employs direct measurements of the average strong-phase differences between $\Dz$ and \Dzb decays within small regions (bins) of the $\D$-decay phase space. Such measurements have been performed by the \besiii and \cleo collaborations using quantum-correlated $\Dz\Dzb$ pairs collected at the $\psiprpr$ resonance~\cite{Ablikim:2020lpk, Ablikim:2020cfp, PhysRevD.82.112006}.  
This model-independent approach results in a small loss in statistical precision due to the binning, in exchange for not relying on modelling the decay amplitudes, resulting in better control of the related systematic uncertainties. The method has been extensively used in \lhcb~\cite{LHCb-PAPER-2020-019, LHCb-PAPER-2018-017, LHCb-PAPER-2014-041}, and the present analysis closely follows the strategies developed therein. This is the first measurement with $B^{\pm}\to \Dstar h^{\pm},$ $\Dstar \to D\piz/\gamma,$ \mbox{$D\to \KS h^+h^-$} decays at \lhcb and the first application of the model-independent method to this channel. The data sample employed in the analysis was collected from proton-proton ($pp$) collisions at centre-of-mass energies of $\sqrt{s}=7,8$ and $13$ \tev, corresponding to a total integrated luminosity of 9 \invfb. 

The paper is organized as follows. The analysis method is described in Sec.~\ref{sec:theory}, while an overview of the \lhcb detector is presented in~Sec.~\ref{sec:Detector}. The data sample selection is summarised in~Sec.~\ref{sec:selection}. The details of the fit procedure are described in~Sec.~\ref{sec:mass_fits}, and the \CP-violating observables measurement in~Sec.~\ref{sec:cpobv}. The systematic uncertainties are reported in Sec.~\ref{sec:systematic}, and the interpretation of the \CP-violating observables in terms of $\ckmgammaangle$ is given in~Sec.~\ref{sec:interpretation}. Finally, conclusions are presented in~Sec.~\ref{sec:conclude}.

%% file: theory.tex
\section{Analysis overview}
\label{sec:theory}

The amplitude for the full decay chain under study is written as 
\begin{equation}
    \label{eq:fullamp}
    \begin{aligned}
        A(\Bm\rightarrow D^{*}h^-) \propto A_D(s_-,s_+) +  f_{\Dstar} A_{\Db}(s_-,s_+) r^{\Dstar h}_Be^{i(\delta^{\Dstar h}_B - \gamma)}, \\
    \end{aligned}
\end{equation}
where $\delta^{\Dstar h}_B$ and $\ckmgammaangle$ are the strong and weak phase differences between $A(B^{-}\rightarrow D^{*0}h^{-})$ (Cabibbo favoured, $b\rightarrow c$) and $A(B^{-}\rightarrow \Dstarzb h^{-})$ (Cabibbo suppressed, $b\rightarrow u$), respectively. The quantity $r^{\Dstar h}_B$ in \cref{eq:fullamp} is the magnitude of the ratio between the two amplitudes and $A_D(s_-,s_+)$ in \cref{eq:fullamp} corresponds to the amplitude of the $\Dz$ decay, 
\begin{equation}
    \begin{aligned}
        A_D(s_-,s_+) \equiv A(\Dz \rightarrow \KS h^+h^-) = |A_D(s_-,s_+)|e^{i\delta_D(s_{-},s_{+})},\\
    \end{aligned}
\end{equation}
where $s_{\mp} \equiv m^2_{\mp}$ is the square of the invariant mass of the $\KS h^{\mp}$ pair. The factor $f_{\Dstar}$ takes the value of $+1$ for the $\Dstar\to\D\piz$ decay and $-1$ for $\Dstar\to\D\gamma$, and derives from the additional \CP-conserving phase shift of $\pi$ between the two decays~\cite{Bondar:2004bi}. The effects of charm mixing and \CP-violation, as well as \CP-violation and matter regeneration in neutral $\KS$ decays, have been studied extensively in the context of such analyses~\cite{Bjorn:2019kov, LHCb-PAPER-2020-019}. They are negligible at the level of precision of this analysis. 
The corresponding \Dzb decay amplitude is thus given by $A_{\Db}(s_-,s_+) = A_D(s_+,s_-)$, and the amplitude for the charge-conjugated $B^+$ decay is obtained with the replacements $A_D(s_-,s_+) \rightarrow A_{D}(s_+,s_-)$ and CKM angle $\gamma \rightarrow -\gamma$.

From \cref{eq:fullamp}, it can be seen that knowledge of $A_D(s_-,s_+)$ is required.  
Bins where $s_- > s_+$ are given a positive index ($+i)$, while the symmetric bin across the  $s_- = s_+$ line is assigned a negative matching index ($-i$). For each bin, the \besiii and \cleo collaborations~\cite{Ablikim:2020lpk, Ablikim:2020cfp, PhysRevD.82.112006} have measured the weighted averages of the cosines ($c_i$) and sines ($s_i$) of the strong phase differences between $\Dz$ and $\Dzb$ decays. Specifically, these quantities are defined as 
\begin{equation}
    \begin{aligned}
        c_i = \frac{\int_i ds_+ ds_-|A_D(s_-,s_+)|^2\cos[\Delta \delta_D(s_-,s_+)]}{\sqrt{\int_i ds_- ds_+ |A_D(s_-,s_+)|^2}\sqrt{\int_i ds_+ ds_- |A_D(s_+,s_-)|^2}},\\
    \end{aligned}
\end{equation}
where the integrals are computed over the $i^{\rm th}$ bin. A similar expression applies for $s_i$ with the cosine replaced by sine. The intensity of the amplitude $|A_D(s_-,s_+)|^{2}$ within a bin is encapsulated by the fraction of (flavour-specific) $\D$ decays occurring in the $i^{\rm th}$ bin, 
\begin{equation}
    \begin{aligned}
        F_i = \frac{1}{N_F}\int_i ds_-ds_+|A_D(s_-,s_+)|^2\eta(s_-,s_+),\\ 
        N_F = \sum_j \int_j ds_-ds_+|A_D(s_-,s_+)|^2\eta(s_-,s_+),
    \end{aligned}
\end{equation}
where $\eta(s_-, s_+)$ corresponds to selection and reconstruction efficiencies. With these definitions, the yields of $\Bm$ and $\Bp$ signal decays in the $i^{\rm th}$ bin are
\begin{equation}
\label{eq:binned_yields}
    \begin{aligned}
        N^{-}_i = H^{-}[F_i+   ({x_-^{\Dstar h}}^2+{y_-^{\Dstar h}}^2)F_{-i}+2f_{D^{*}}\sqrt{F_i F_{-i}}(c_i x^{\Dstar h}_-+s_i y^{\Dstar h}_-)], \\
        N^{+}_i = H^{+}[F_{-i}+({x_+^{\Dstar h}}^2+{y^{\Dstar h}_+}^2)F_{i}+ 2f_{D^{*}}\sqrt{F_i F_{-i}}(c_i x^{\Dstar h}_+-s_i y^{\Dstar h}_+)], \\
    \end{aligned}
\end{equation}
respectively. The quantities $x^{\Dstar h}_{\pm}, y^{\Dstar h}_{\pm}$ are the Cartesian \CP-violating observables. The normalisation factors $H^{+}$ and $H^{-}$ are different for each decay and each flavour due to different production and detection asymmetries.

The Cartesian \CP-violating observables are written in terms of the weak phase $\ckmgammaangle$ and hadronic parameters $r_B^{\Dstar h}, \delta_B^{\Dstar h}$,
\begin{equation}
    \begin{aligned}
        x_{\pm}^{\Dstar h} = r_B^{\Dstar h} \cos(\delta_B^{\Dstar h} \pm \gamma),\\
        y_{\pm}^{\Dstar h} = r_B^{\Dstar h} \sin(\delta_B^{\Dstar h} \pm \gamma).
    \end{aligned}
\end{equation} 
 
The $\Bpm\rightarrow \Dstar \pipm$ and $\Bpm\rightarrow \Dstar \Kpm$ decays have their own set of $x_\pm$ and $y_\pm$ parameters, respectively, giving a total of 8 \CP-violating observables. It is convenient to reparameterise the \CP-violating observables for the $\Bpm\rightarrow \Dstar \pipm$ decay~\cite{LHCb-PAPER-2020-019, GarraTico:2018nng, PhysRevD.102.053003}, 
\begin{equation}
\xi^{\Dstar  \pi} = (r^{\Dstar \pi}_{B}/r^{\Dstar  K}_{B})\exp[ i(\delta_B^{\Dstar \pi} - \delta_{B}^{\Dstar  K})]
\label{eq:cpxi}
\end{equation}
with $x_{\xi}^{\Dstar \pi} \equiv \text{Re}(\xi^{\Dstar \pi})$ and $y_{\xi}^{\Dstar \pi} \equiv \text{Im}(\xi^{\Dstar \pi})$. With this reparameterisation, the \CP-violating observables are expressed as
\begin{equation}
    x^{\Dstar  \pi}_{\pm} = x^{\Dstar  \pi}_{\xi}x^{\Dstar  K}_{\pm} - y^{\Dstar  \pi}_{\xi}y^{\Dstar  K}_{\pm}, \qquad y^{\Dstar  \pi}_{\pm} = x^{\Dstar  \pi}_{\xi}y^{\Dstar  K}_{\pm} + y^{\Dstar  \pi}_{\xi}x^{\Dstar  K}_{\pm}.
    \label{eq:cpdstpi}
\end{equation}
This parametrisation utilises the fact that $\gamma$ is common in the definition of $x^{D^*h}_{\pm}$ and $y^{D^*h}_{\pm}$ to reduce the number of fit parameters, and thus makes the fit more stable~\cite{LHCb-PAPER-2020-019, GarraTico:2018nng, PhysRevD.102.053003}.

The $F_i$ parameters are used for both the $\Bpm\to \Dstar \pipm$ and $\Bpm\to \Dstar  \Kpm$ channels, though due to the larger sample size, it is primarily the $\Bpm\rightarrow \Dstar \pipm$ channel that determines them. For this approach to be valid, it is necessary that the efficiencies be consistent among the two channels over the phase space. This is a reasonable assumption due to the identical topologies and nearly identical candidate selections, with the only differences being related to identifying the hadron produced in the $\Bpm$ decay. Additionally, there are specific background components associated with the $D$ decay. The yields of these backgrounds can be expressed in terms of the quantities $F_i$, thereby reducing the number of fit parameters.

\begin{figure}
\centering

\includegraphics[width=0.48\textwidth]{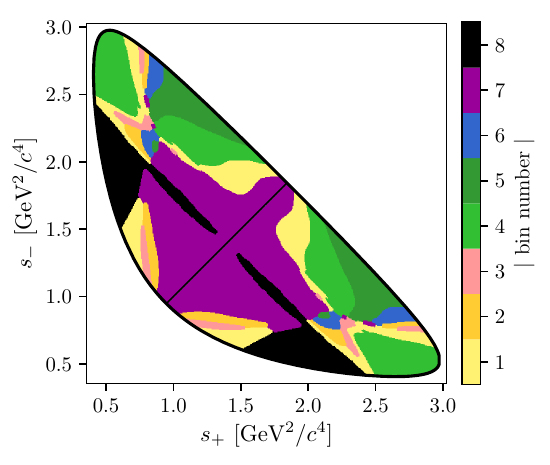}
\includegraphics[width=0.48\textwidth]{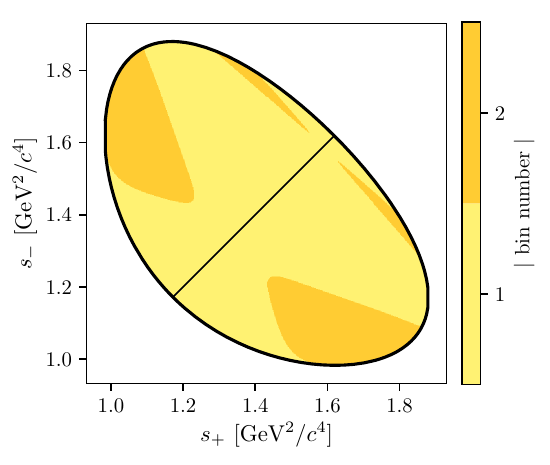}
\caption{(left) ``Optimal'' binning scheme~\cite{Ablikim:2020lpk} used for $D\rightarrow \KS \pi^+\pi^-$ decays. (right) ``2-bins'' binning scheme~\cite{Ablikim:2020cfp} used for $D\rightarrow \KS K^+K^-$ decays. They are the same as those in a previous model-independent measurement~\cite{LHCb-PAPER-2020-019}. }
\label{fig:binning}. 
\end{figure}

In this analysis, the ``optimal'' binning scheme~\cite{Ablikim:2020lpk, PhysRevD.82.112006} of the $D\rightarrow \KS \pi^+\pi^-$ decays, which takes into account the model of $\Dz$ decays and distribution of $B^{\pm}$ decays and improves sensitivity to $\ckmgammaangle$, is employed, and is shown in \cref{fig:binning}. 
Also shown is the `2-bins' binning scheme used for the $D\rightarrow \KS K^+K^-$ decays~\cite{Ablikim:2020cfp}, where fewer bins are used due to the lower signal yields. 

%% file: detector.tex
\section{Detector, simulation and analysis}
\label{sec:Detector}

The \lhcb detector~\cite{LHCb-DP-2008-001,LHCb-DP-2014-002} is a single-arm forward
spectrometer covering the \mbox{pseudorapidity} range $2<\eta <5$, primarily
designed for the study of particles containing \bquark or \cquark
quarks. The detector includes a high-precision tracking system
consisting of a silicon-strip vertex detector surrounding the $pp$
interaction region~\cite{LHCb-DP-2014-001}, a large-area silicon-strip detector located
upstream of a dipole magnet with a bending power of about
$4{\mathrm{\,Tm}}$, and three stations of silicon-strip detectors and straw
drift tubes~\cite{LHCb-DP-2013-003,LHCb-DP-2017-001}
placed downstream of the magnet.
The tracking system provides a measurement of the momentum, \ptot, of charged particles with
a relative uncertainty that varies from 0.5\% at low momentum to 1.0\% at 200\gevc.
The minimum distance of a track to a primary $pp$ collision vertex (PV), the impact parameter (IP), 
is measured with a resolution of $(15+29/\pt)\mum$,
where \pt is the component of the momentum transverse to the beam, in\,\gevc.
Different types of charged hadrons are distinguished using information
from two ring-imaging Cherenkov detectors~\cite{LHCb-DP-2012-003}. 
Photons, electrons and hadrons are identified by a calorimeter system consisting of
a scintillating pad and pre-shower detectors, an electromagnetic 
and a hadronic calorimeter. Muons are identified by a system composed of alternating layers of iron and multiwire
proportional chambers~\cite{LHCb-DP-2012-002}.
The online candidate selection is performed by a trigger~\cite{LHCb-DP-2012-004}, 
which consists of a hardware stage, based on information from the calorimeter and muon
systems, followed by a software stage, which applies a full candidate reconstruction.

The candidates to be reconstructed and considered in this analysis are triggered at the hardware stage, based on information from the calorimeter and muon systems, when one of the final-state particles in the signal decays has deposited high transverse energy in the calorimeter (trigger on the signal events) or one of the other particles not in the signal decays fulfils any trigger requirement (trigger independent of the signal events). For the software-stage trigger, which applies a full event reconstruction, at least one particle is required to have high \pt and $\chisqip$, defined as the difference in the vertex-fit $\chisq$ of the PV reconstructed with and without the particle under consideration. 
A multivariate algorithm~\cite{BBDT} is used for the identification of secondary vertices consistent with the two-track, three-track or four-track decay of a \bquark hadron. The PVs are fitted with and without the signal $B$ tracks, and the PV with the smallest value of $\chi^2_{\rm IP}$ is associated with the $B$ meson candidate.

Simulation samples are used to determine the shapes of the invariant mass distributions of the reconstructed $DK^{\pm} (\pi^{\pm})$ and $D\pi^0 (\gamma)$ candidates for both the signal and background processes. They are also used to validate assumptions about the relative efficiency variations of the different contributing processes. In the simulation, $pp$ collisions are generated using
  \pythia~\cite{Sjostrand:2007gs,*Sjostrand:2006za} 
  with a specific \lhcb configuration~\cite{LHCb-PROC-2010-056}.
  Decays of unstable particles
  are described by \evtgen~\cite{Lange:2001uf}, in which final-state
  radiation is generated using \photos~\cite{davidson2015photos}. 
  The interaction of the generated particles with the detector, and its response,
  are implemented using the \geant
  toolkit~\cite{Allison:2006ve, *Agostinelli:2002hh} as described in
  Ref.~\cite{LHCb-PROC-2011-006}. 
  The $D\rightarrow \KS \pi^+\pi^-$ and $D\rightarrow \KS K^+K^-$ decays are generated uniformly in the $D$ decay phase space. Fast simulations~\cite{Cowan:2016tnm}, which consider the geometric acceptance and tracking efficiency of \lhcb as well as the dynamics of the decay, are also used to model background lineshapes.

%% file: selection.tex
\section{Signal candidate selection criteria}
\label{sec:selection}

The selection criteria of the charged particles closely follow those in the previous \lhcb measurement~\cite{LHCb-PAPER-2020-019}, with the main differences being due to the reconstruction of the additional $\piz$ meson or photon. 
For the final-state charged tracks used to form combinations of intermediate particles, requirements are placed on the track quality, momenta, and IP to suppress random tracks coming from the PV. 

The $\KS$ meson candidate is reconstructed using two oppositely charged tracks identified as pions.  
To form a $\D$ meson candidate, the $\KS$ candidate is combined with two oppositely charged tracks, both assigned with either a kaon or pion hypothesis. The photons are reconstructed by the formation of energy clusters in the calorimeter system which are not associated with reconstructed tracks. The $\piz$ mesons are reconstructed by combining pairs of energy clusters in the calorimeter system that are consistent with the photon hypothesis (two photons in two different cells of the electromagnetic calorimeter). The $\Dstar $ candidates are reconstructed by combining a $\D$ candidate with a reconstructed $\piz$ (for the decay $\Dstar \rightarrow \D \piz$) or a photon (for the decay $\Dstar \rightarrow \D \gamma$). Finally, the $\Bpm$ candidates are formed by combining a $\Dstar $ candidate with a final track, identified either as a pion or a kaon. 

Particle identification (PID) requirements are used to separate the reconstructed \mbox{$\Bpm\to\Dstar \Kpm$} and \mbox{$\Bpm\to\Dstar \pipm$} samples into disjoint sets. The misidentification rates are approximately 13\% and 3\% for the $\Bpm \rightarrow \Dstar \Kpm$ and $\Bpm \rightarrow \Dstar \pipm$ decays, respectively. Loose PID requirements are also applied to the $\D$ decay products to reduce the backgrounds due to random combinations of tracks. The charged hadron tracks produced in the $\Bpm$ and $\D$ decays are required to have no corresponding activity in the muon detector to suppress muonic decays. Further PID requirements are imposed on the $\D$ products to reject decays to final states with electrons.
The significance of the distance between the $\D$ decay vertex and the \Bpm decay vertex in the beam direction is used to remove background that decays to the same final state but does not proceed through the intermediate $\Dz$ meson, \ie $\Bpm\to \KS h^{'+} h^{'-} h^{\pm}$. 
Similarly, to suppress backgrounds from $\D\to \pip\pim\pip\pim$ and $\D\to \pip\pim \Kp\Km$, the \KS vertex is required to be significantly displaced from the $\D$ vertex. 

Multivariate algorithms~\cite{Hocker:2007ht, TMVA4} are applied to reduce the combinatorial background. Two boosted decision trees~\cite{Breiman, AdaBoost} (BDT classifiers) are employed. The first BDT classifier focuses on the charged final-state tracks and is referred to as the charged BDT classifier. This classifier is the same as used in a previous \lhcb $\Bpm\to D\hpm$ analysis~\cite{LHCb-PAPER-2020-019}, where the charged candidates have very similar kinematics. The variables used to identify signal-like candidates include the momenta of the $\Bpm$, $\D$, and companion particles, the vertex positions, and the parameters that describe the fit quality in the reconstruction. The BDT classifier was trained on simulated signal samples and the upper sideband of the $\Bpm$ invariant mass as a background proxy. The charged BDT classifier requirement providing the best sensitivity to $\ckmanglegamma$ is determined based on a series of pseudo-experiments.
An additional BDT classifier is used to reduce the combinatorial background associated with the $\Dstar\to\D \piz/\gamma$ reconstruction and is referred to as the neutral BDT classifier. Separate classifiers are trained for the $\Dstar\to\D\gamma$ and $\Dstar \to \D\piz$ samples. The momentum of the $\piz/\gamma$ as well as the confidence level of a cluster in the calorimeters produced by a $\gamma$~\cite{LHCb-DP-2020-001}, which discriminates against clusters associated with charged hadrons, are used as discriminating variables. The simulation samples are used as signal proxies, while the background training sample is obtained from the $D + \pi^0 (\gamma)$ invariant-mass range of 2035-2100 (2065-2160) MeV/$c^2$. The working points of the neutral BDT classifiers are chosen separately with pseudoexperiments to optimize sensitivity to the CKM angle $\gamma$, besides the above optimisation for the charged BDT classifier. The optimisation is performed individually for the $\pi^0$ and $\gamma$ modes. The charged and neutral BDT requirements together have efficiencies of approximately 62\% and 78\% for the $\Bpm \rightarrow [\D \piz]_{\Dstar} \Kpm$ and $\Bpm \rightarrow [\Dstar\rightarrow \D \gamma]_{\Dstar} \Kpm$ decays, respectively.

In the above selections, the invariant masses of $D$, $\KS$ and $\pi^0$ mesons are constrained to their known masses~\cite{PDG2022} and the $B^{\pm}$ momentum vectors are required to point to the PV, whenever applicable, to improve the mass resolution.

%% file: mass_fits.tex
\section{Invariant mass distribution parameterisations}
\label{sec:mass_fits}

 Two-dimensional fits to the $m(D\pi^0/\gamma)$ and $m(Dh^{\pm})$ invariant masses are performed to distinguish signal from backgrounds, where $h^{\pm}$ is the hadron from the $B^{\pm}$ meson decays. It is essential for one of these variables to be $\mdn$, to separate correctly reconstructed \mbox{${\Dstar \to \D\piz/\gamma}$} signal contributions from two types of backgrounds in the $D^\ast$ reconstruction. Random combinations of unrelated $\D$ and neutral particles can be incorrectly reconstructed as a $D^\ast$ (in-reco $D^\ast$). A single photon from $\piz\to \gamma\gamma$ combined with the correct $\D$ leads to a partially reconstructed $D^\ast$ (part-reco $D^\ast$) in the $D^\ast \to D\gamma$ mode. These partially and incorrectly reconstructed decays have sensitivity to $\gamma$ and are thus accounted for, in the mass fits. The second variable is chosen to be $m(Dh^{\pm})$, instead of $m(\Dstar h^{\pm})$, to avoid large correlations with \mdn. The $m(Dh^{\pm})$ distribution allows for distinguishing signal, partially reconstructed and misidentified $\Bpm$ decays, and combinatorial background (random combinations of a $D$ candidate with a hadron). 
The components entering the fit and their corresponding probability density functions (PDF) are described henceforth.

The shape parameters are obtained from either simulation samples of the signal and background processes or from data-driven methods that utilise the random combinations of reconstructed particles in data samples to mimic the combinatorial behaviour of backgrounds. The shapes are considered to be the same for all the $D$ Dalitz plot (DP) bins and any possible differences are accommodated in the systematic uncertainties.

The parameterisation of the \mdn distribution assumes that all true $\Dstar$ decays present in the sample have the same shape, regardless of whether they came from a signal $\Bpm\to\Dstar\hpm$ decay or from a partially reconstructed $B$ background decay. This assumption has been validated using simulated samples. The incorrectly reconstructed $\Dstar$ candidates are predominantly combinatorial in nature and are modelled by randomly combining $\D$ and $\piz/\gamma$ particles from different candidates and smoothing the resulting distribution with an empirical function. The empirical function comprises different polynomial functions in various mass ranges.

 The $\D\piz$ invariant-mass distribution of $\Dstar\to \D\piz$ signal decays is described with a sum of a Gaussian function and a Crystal Ball (CB) function~\cite{Skwarnicki:1986xj}. The two components share a common mean, which is allowed to vary freely in the fit. 
In order to account for differences in resolution between the data and simulation, the shapes determined from simulations are convoluted with a Gaussian function whose resolution is determined in the fit. 

The $\D\gamma$ invariant-mass distribution of $\Dstar\to \D\gamma$ signal decays is modelled with a single CB function. The mean and resolution of the CB function are allowed to vary freely in the fit while other parameters are fixed using simulation samples. There is an additional contribution for the $m(\D \gamma)$ fit due to partially reconstructed $\Dstar(\to \D\piz)$ decays, in which only one $\gamma$ from the $\piz$ decay is reconstructed. This contribution is modelled by a polynomial convolved with a CB function. The difference of the mean between the data and simulation is fixed to that found in the $\Dstar\rightarrow \D\gamma$ decays. The resolution of the CB function is allowed to vary freely, while the rest of the parameters are fixed according to the simulation. 
If $B^{\pm}$ candidates are reconstructed in both $\Dstar\rightarrow \D\piz$ and $\Dstar\rightarrow \D\gamma$ final states in the same event, the candidate reconstructed with $\Dstar\rightarrow \D\gamma$ is removed in order to prevent double counting due to these partially reconstructed $\Dstar$ contributions. 

The \mdh distributions of $\Bpm\to\Dstar h^{\pm}$ decays have been extensively studied in previous analyses~\cite{LHCb-PAPER-2020-036, LHCb-PAPER-2017-021}, and the parameterisations developed therein are used to describe the contributions here. The angular distributions of the $\Dstar$ decay drive the expected \mdh shape and depend on the spin of the $\piz/\gamma$ in the decay. In particular, $\Dstar\to\D\piz$ results in a double-peaking structure in the invariant-mass distribution range of 5000--5200 \mevcc. It is described by a parabolic function to model the effect of the angular distribution, multiplied by a linear function to take into account efficiency effects, and convolved with a double Gaussian function to incorporate resolution effects~\cite{LHCb-PAPER-2017-021}. Meanwhile, $\Dstar\to\D\gamma$ results in a hill structure around 4950--5200 \mevcc, and is described by an inverted parabola, multiplied by another linear function, and convolved with a double Gaussian function~\cite{LHCb-PAPER-2017-021}.

There is a clear contribution from $B^{\pm}\to \D \hpm$ decays that have been combined with a random neutral particle. The contribution is modelled as the sum of a Gaussian function and a modified Cruijff function~\cite{LHCb-PAPER-2020-019} and is considered as combinatorial in the \mdn distribution.  
The misidentified and partially reconstructed backgrounds are described with empirical functions determined from simulation.  
These backgrounds include \mbox{$B^{0,\pm}\rightarrow \D\pi^{\pm}\pi^{0,\mp}$},\mbox{ $B^{0,\pm}\rightarrow \D K^{\mp}\pi^{\pm,0}$}, \mbox{$B^0_s\rightarrow \D K^{\pm}\pi^{\mp}$}, \mbox{$B^{0,\pm}\rightarrow \Dstar \hpm\pi^{\mp,0}$} and \mbox{$B_s^0\rightarrow \D^{*} K^{\pm} \pi^{\mp}$} decays, where possible intermediate resonant contributions in the $B$ decay are included.

Finally, an exponential function is used to describe the small contribution from combinatorial $\D$ and $\hpm$ pairs. The slope of the exponential function is a free parameter in the fit.

%% file: cpobv.tex
\section{ \texorpdfstring{\CP}{CP}-violating observables }
\label{sec:cpobv}
\begin{figure}
    \centering
    \hfill
    \subfigure[]{
    \centering
    \includegraphics[width=0.25\textwidth]{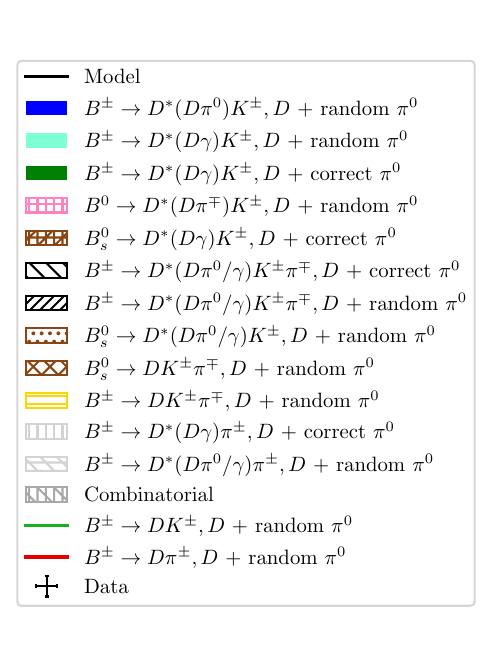}
    \includegraphics[width=0.74\textwidth]{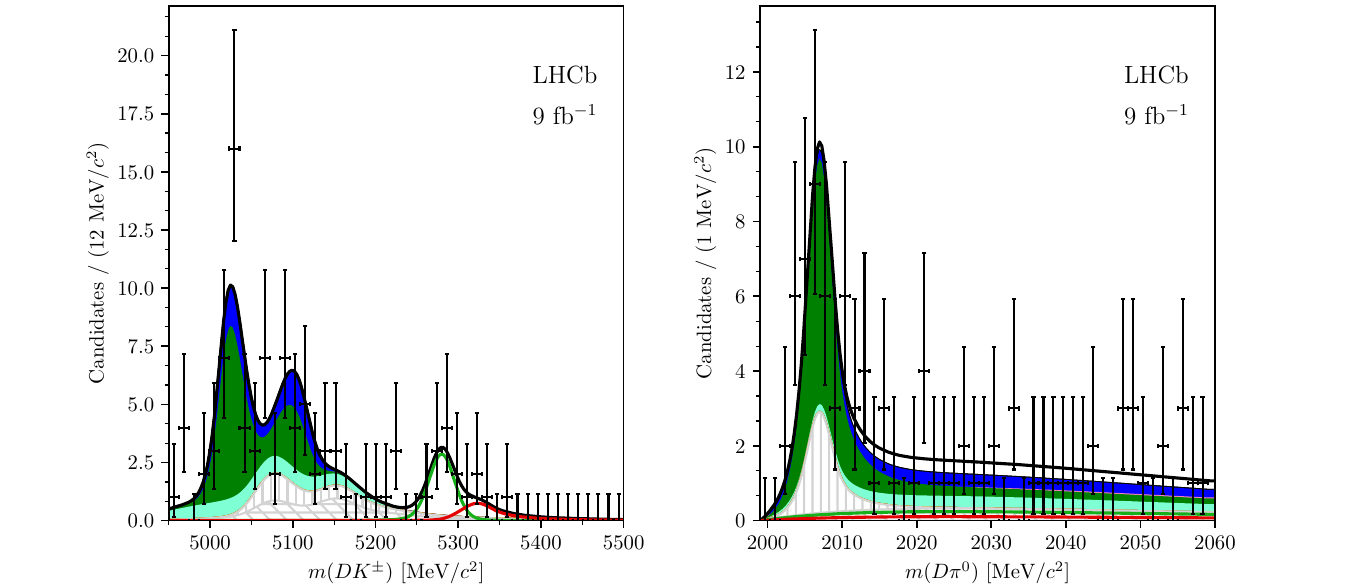}
    \label{fig:dataset_pi0_d2kskk_dk}}
    \hfill
    \subfigure[]{
    \centering
    \includegraphics[width=0.25\textwidth]{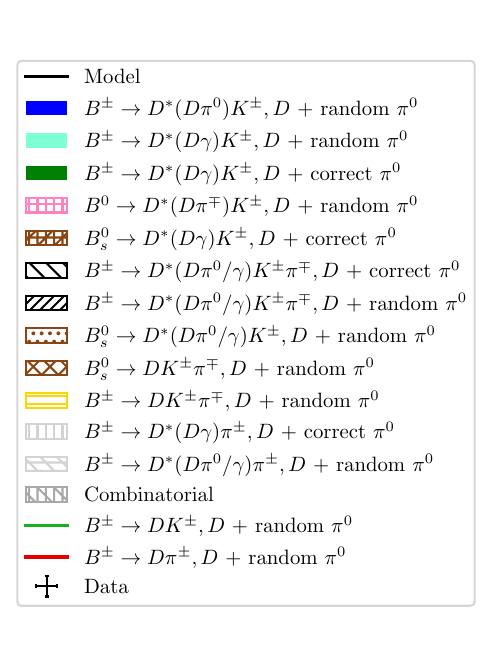}
    \includegraphics[width=0.74\textwidth]{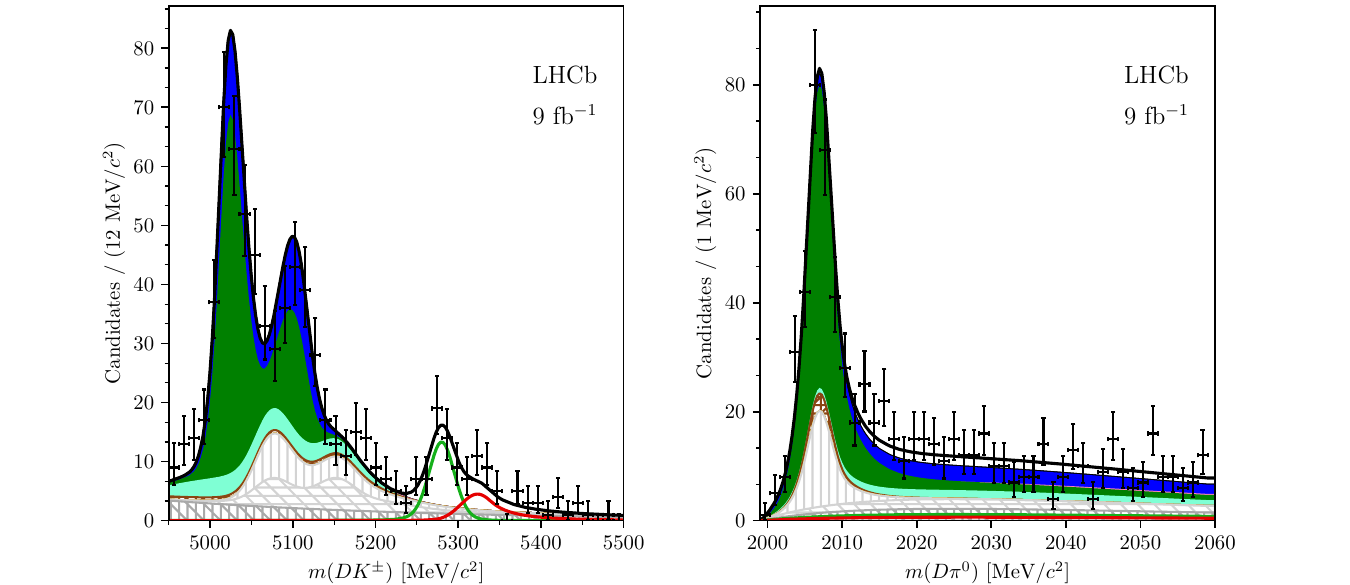}
    \label{fig:dataset_pi0_d2kspipi_dk}
    }
    \caption{Projections of (left) $m(\D \hpm)$ invariant-mass distribution and (right) $m(\D\piz)$ invariant-mass distribution for the $\Bpm\to\Dstar \Kpm$ channels, reconstructed with the \mbox{$\Dstar\to \D\piz$}, (a) \mbox{$\D\to \KS \Kp\Km$} and (b) \mbox{$\D\to \KS \pi^+\pi^-$} modes. The signal and in-reco $\Dstar$ contributions are shown as solid colours.}
    \label{fig:dataset_pi0_dk}
\end{figure}

\begin{figure}
    \centering
    \hfill
    \subfigure[]{
    \centering
    \includegraphics[width=0.25\textwidth]{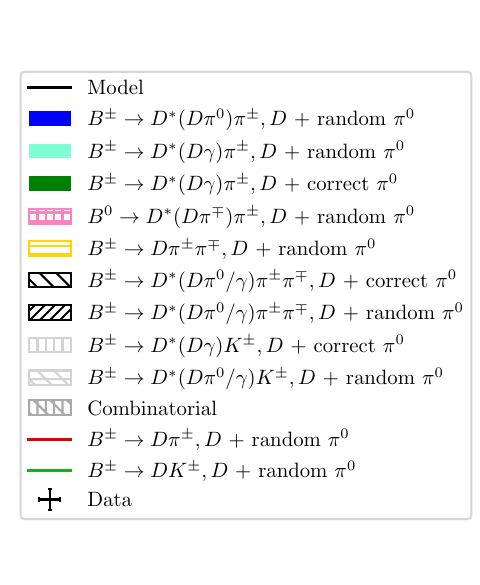}
    \includegraphics[width=0.74\textwidth]{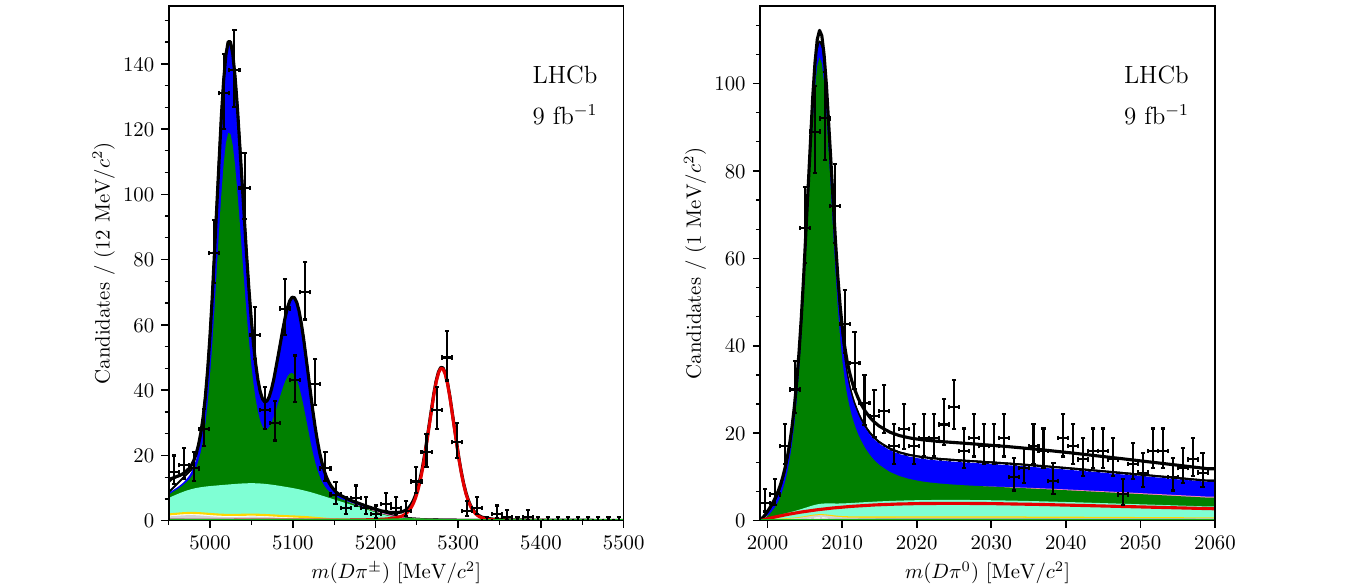}}
    \hfill
    \subfigure[]{
    \centering
    \includegraphics[width=0.25\textwidth]{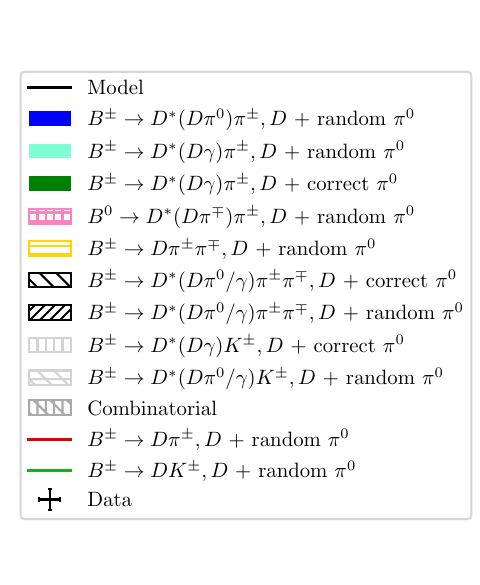}
    \includegraphics[width=0.74\textwidth]{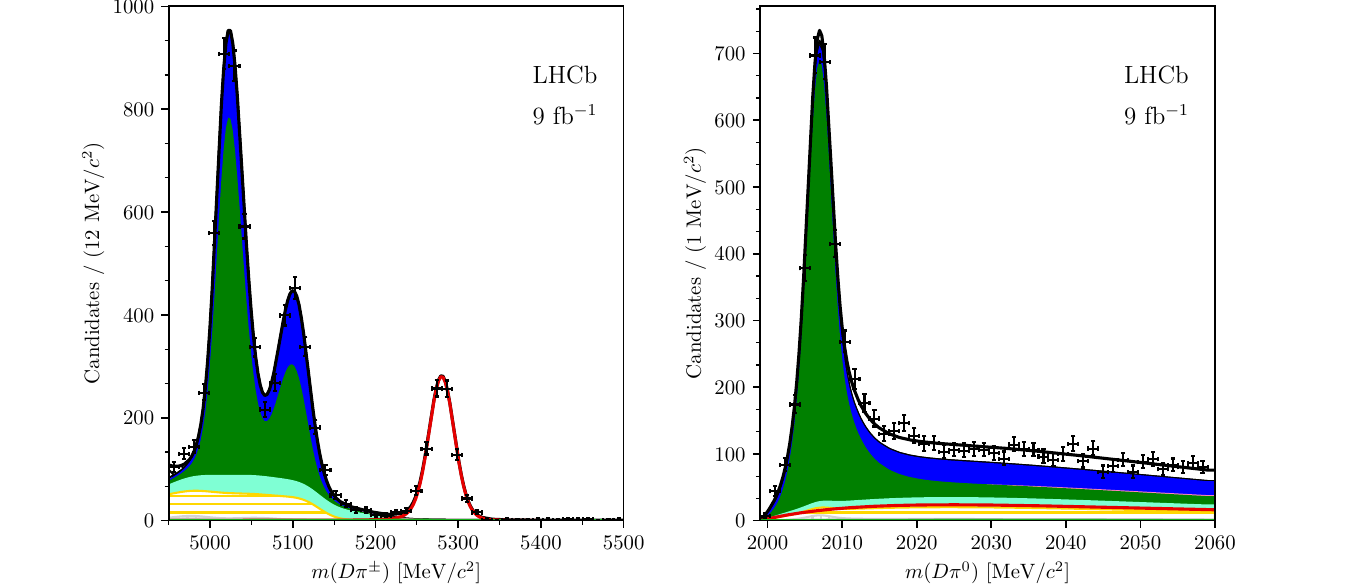}}
    \caption{Projections of (left) $m(\D \hpm)$ invariant-mass distribution and (right) $m(\D\piz)$ invariant-mass distribution for the $\Bpm\to\Dstar \pipm$ channels, reconstructed with the $\Dstar\to \D\piz$, (a) $\D\to \KS \Kp\Km$ and (b) $\D\to \KS \pi^+\pi^-$ modes. The signal and in-reco $\Dstar$ contributions are shown as solid colours.}
    \label{fig:dataset_pi0_dpi}
\end{figure}

\begin{figure}
    \centering
    \hfill
    \subfigure[]{
    \centering
    \includegraphics[width=0.25\textwidth]{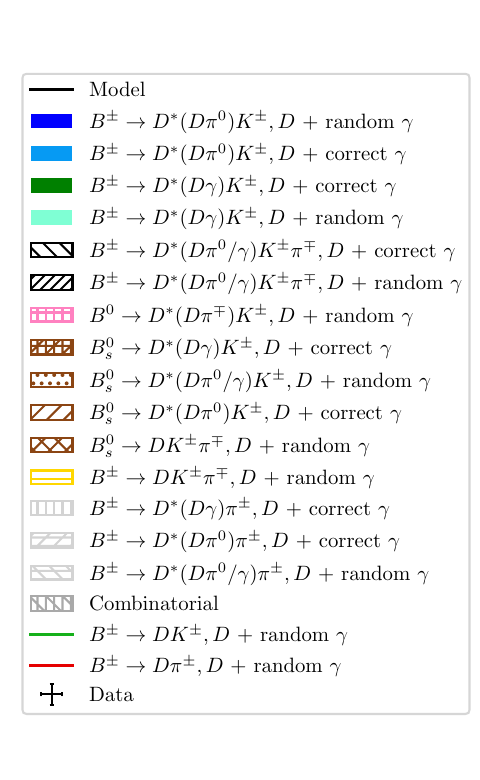}
    \includegraphics[width=0.74\textwidth]{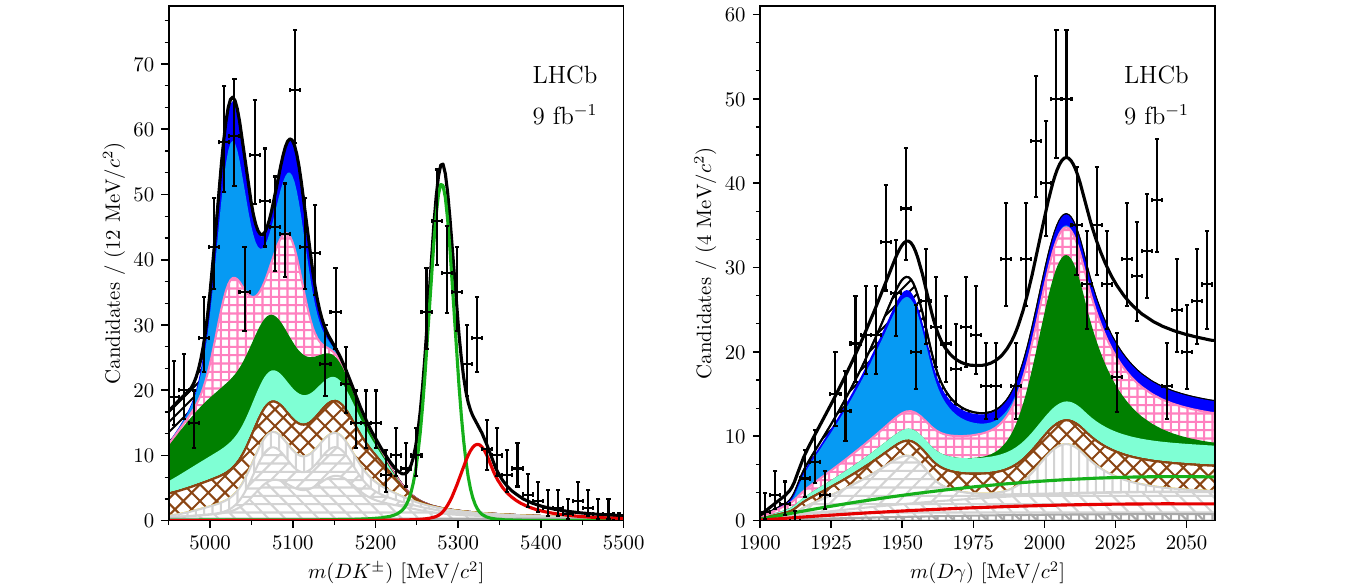}}
    \hfill
    \subfigure[]{
    \centering
    \includegraphics[width=0.25\textwidth]{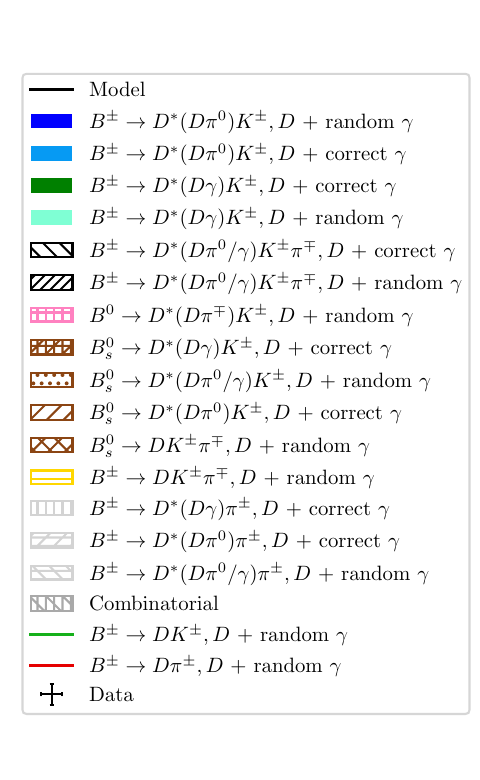}
    \includegraphics[width=0.74\textwidth]{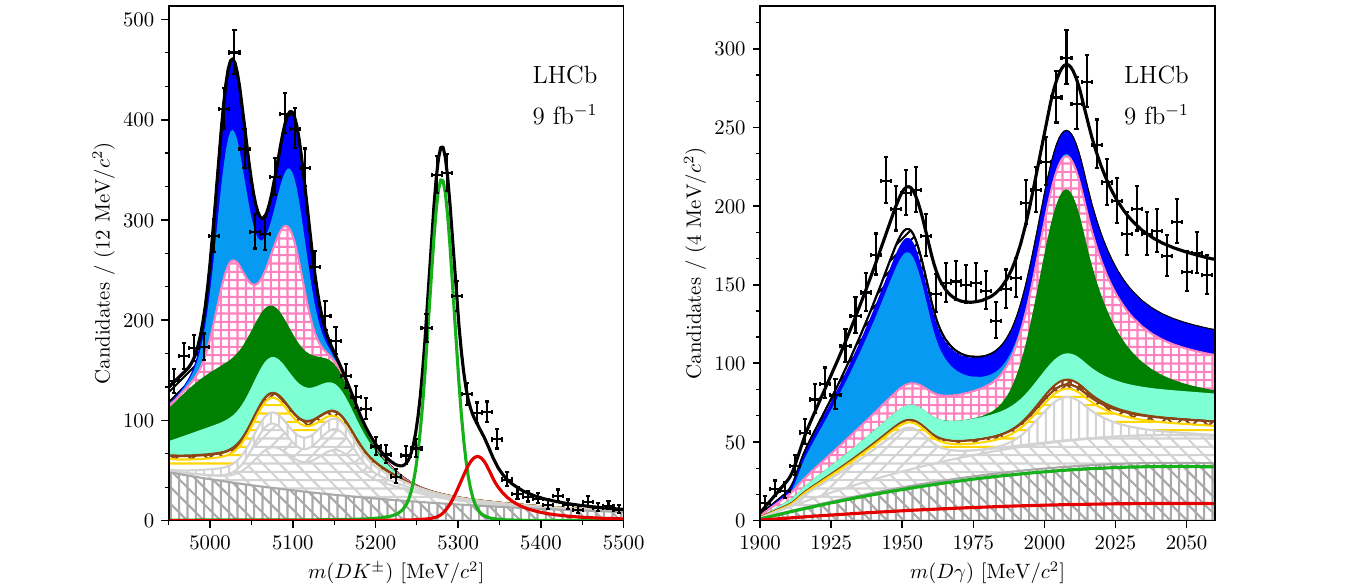}}
    \caption{Projections of (left) $m(\D \hpm)$ invariant-mass distribution and (right) $m(\D\gamma)$ invariant-mass distribution for the $\Bpm\to\Dstar \Kpm$ channels, reconstructed with the $\Dstar\to \D\gamma$, (a) \mbox{$\D\to \KS \Kp\Km$} and (b) $\D\to \KS \pi^+\pi^-$ modes. The signal, in-reco $\Dstar$ and part-reco \Dstar contributions are shown as solid colours.}
    \label{fig:dataset_gamma_dk}
\end{figure}

\begin{figure}
    \centering
    
    \subfigure[]{
    \centering
    \includegraphics[width=0.25\textwidth]{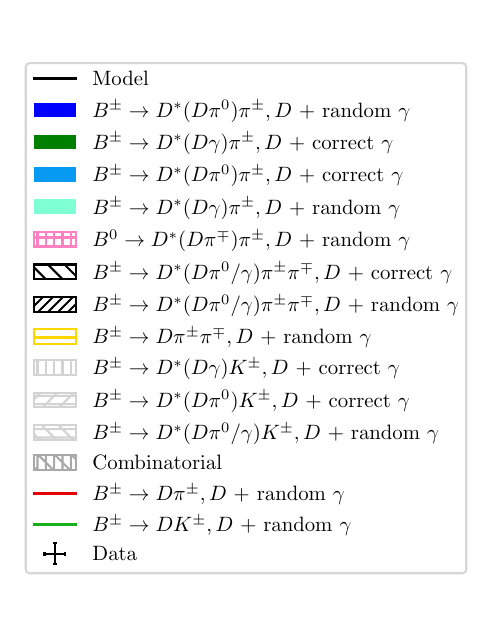}
    \includegraphics[width=0.74\textwidth]{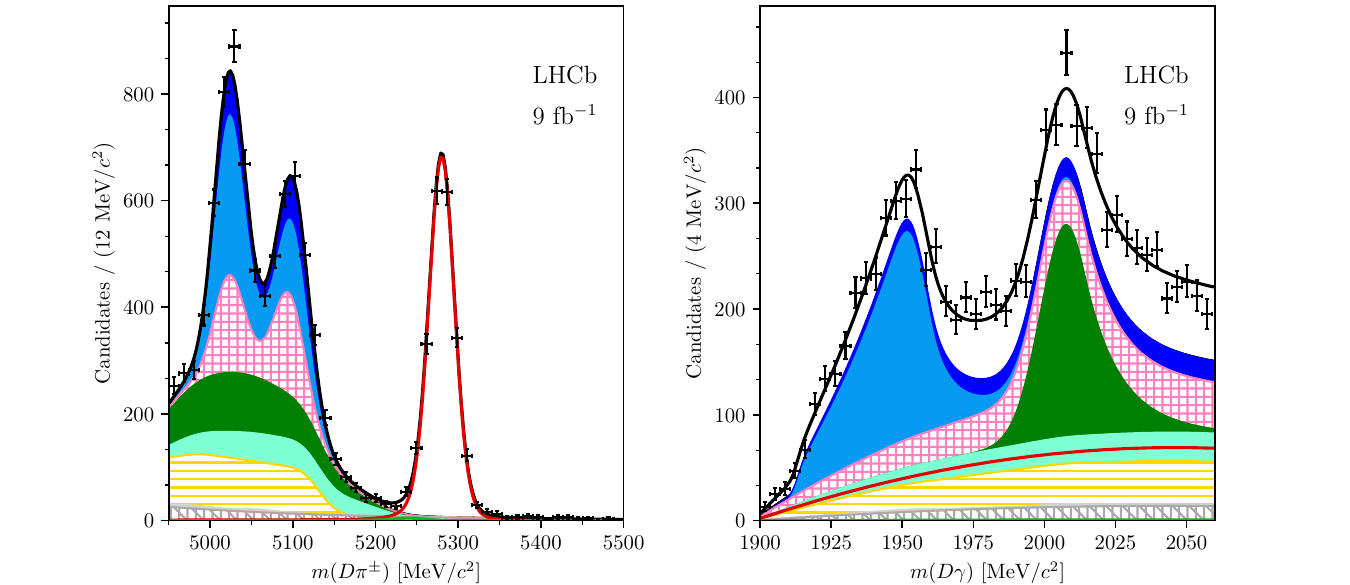}}

    \subfigure[]{
    \centering
    \includegraphics[width=0.25\textwidth]{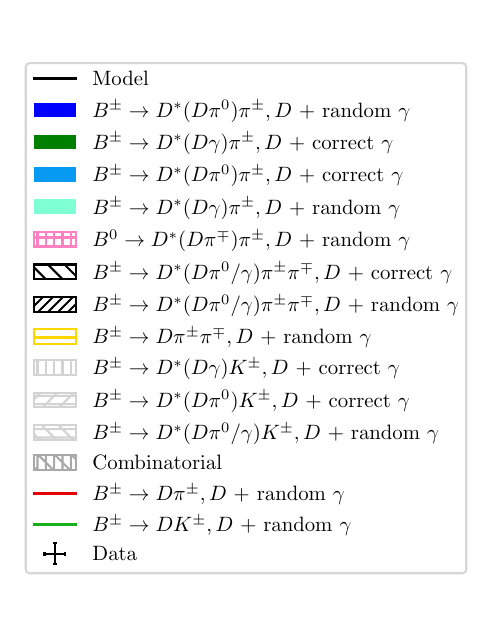}
    \includegraphics[width=0.74\textwidth]{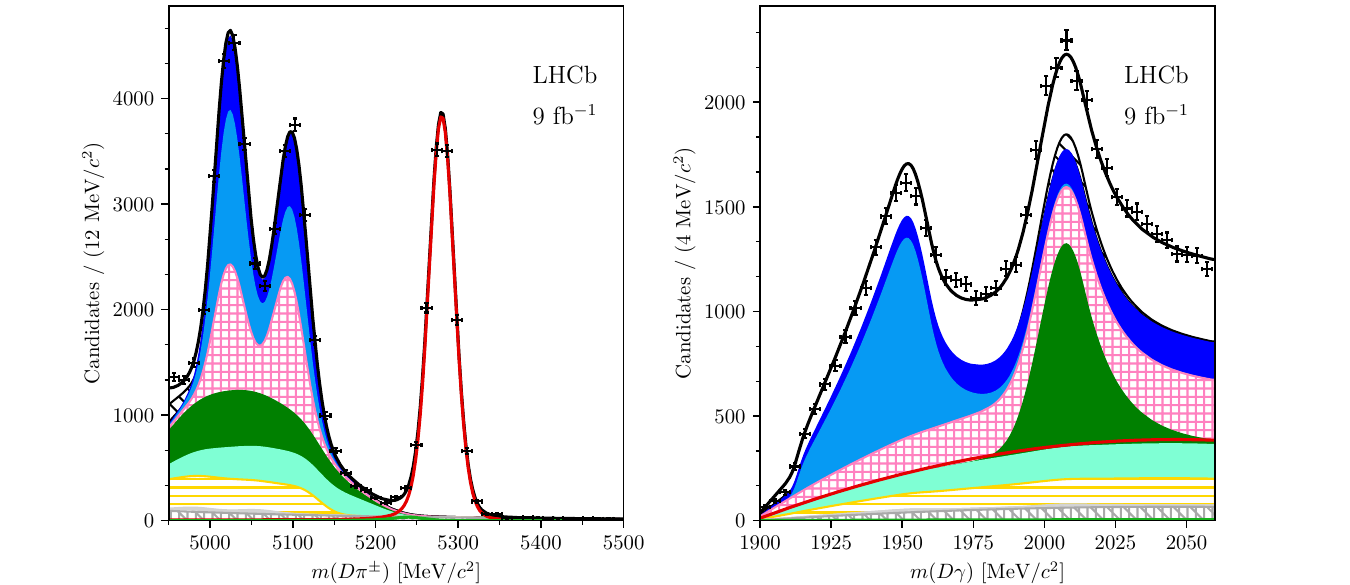}}
    \caption{Projections of (left) $m(\D \hpm)$ invariant-mass distribution and (right) $m(\D\gamma)$ invariant-mass distribution for the $\Bpm\to\Dstar \pipm$ channels, reconstructed with the $\Dstar\to \D\gamma$, (a) \mbox{$\D\to \KS \Kp\Km$} and (b) \mbox{$\D\to \KS \pi^+\pi^-$} modes. The signal, in-reco $\Dstar$ and part-reco \Dstar contributions are shown as solid colours.}
    \label{fig:dataset_gamma_dpi}
\end{figure}

The data sample is split according to the charge of the parent $\B$ meson, $\Bpm$ decay ($\Bpm\rightarrow\Dstar\pi^{\pm}, \Bpm\rightarrow\Dstar K^{\pm}$), $\Dstar$ decay ($\Dstar\rightarrow D\piz, \Dstar\rightarrow D\gamma$), $D$ decay (\mbox{$\D\rightarrow \KS K^+K^-, \D\rightarrow \KS \pi^+\pi^-$}), and DP bins (16 for $\KS \pi^+\pi^-$ and 4 for $\KS K^+K^-$). There are 320 categories in total, 256 for the $\KS \pi^+\pi^-$ modes and 64 for the $\KS K^+K^-$ modes. 
An unbinned extended maximum-likelihood fit is performed simultaneously to the 2D invariant-mass distributions (\mdh and \mdn in the mass range 4950-6000 \mevcc and 1999-2010/1900-2060 \mevcc) for all the above categories to extract the \CP-violating observables.  
The results of the invariant-mass fits are shown in Figs.~\ref{fig:dataset_pi0_dk} and \ref{fig:dataset_pi0_dpi} for the $\Dstar\to \D\piz$ modes, and in Figs.~\ref{fig:dataset_gamma_dk} and \ref{fig:dataset_gamma_dpi} for the \mbox{$\Dstar\to \D\gamma$} modes. In these figures, the top plots correspond to the $\D\to\KS K^+K^-$ modes and the bottom plots correspond to the $\D\to\KS \pi^+\pi^-$ modes. The different contributions are shown in the left legends, which give the considered decay channels and its reconstructed categories.  
The mass shapes and parameters for various physics contributions were introduced in Sec.~\ref{sec:mass_fits}. 


In $B^{\pm}\to \Dstar h^{\pm}$ signal decays and $B^{\pm}\to \D \hpm$ background contributions, 
the yields in different DP bins are allotted according to \cref{eq:binned_yields}.  
The \CP-violating observables $x^{D^{(*)}h}_{\pm}, y^{D^{(*)}h}_{\pm}$, $x^{D^{(*)}h}_{\xi}$ and $y^{D^{(*)}h}_{\xi}$ for $B^{\pm}\to \Dstar h^{\pm}$ signal and $B^{\pm}\to \D \hpm$ decays are varied freely.  
The $c_i, s_i$ parameters are external inputs, fixed to the values measured from the \besiii and \cleo collaborations~\cite{Ablikim:2020lpk, Ablikim:2020cfp, PhysRevD.82.112006}. To reduce fit instabilities due to the strong correlations among the $F_i$ variables, they are reparameterized as a series of recursive fractions, $R_i$, as in the $B^{\pm}\to Dh^{\pm}$ measurement~\cite{LHCb-PAPER-2020-019}. The relations are given by 
\begin{equation}
     F_i = \left\{
     \begin{aligned}
         R_i,\ \ \ \ \ \ \ \ \ \ \ \ \ \ \  &  & i=-N , \\
         R_i \prod_{j<i}(1-R_j) & , & -N < i < +N , \\
         \prod_{j<i}(1-R_j) & , & i = +N. 
     \end{aligned}
     \right.
 \end{equation}
The $R_i$ parameters are varied freely in the fit. The $R_i$ parameters are shared between all the components as the other hadrons directly produced in the $B$ decays are found to have little impact on the relative efficiency across the DP plane.

For the $B^{0,\pm}\rightarrow \D \pi^{\pm}\pi^{\mp,0}$ decay mode, the yields in different DP bins are varied freely as \CP-violation effects may be present and the yields are sufficiently large. The combinatorial background yields are also varied freely separately for each bin, as charge-asymmetric effects may be present and the distribution across the phase space cannot be predicted reliably. \CP symmetry is assumed for other background modes, as the yields are too low to be sensitive to \CP-violation effects or such effects are expected to be sufficiently small. This is the case for the \mbox{$B^{0,\pm}\rightarrow \D K^{\pm}\pi^{\mp,0}$}, \mbox{$B_s^{0}\rightarrow \D^{(*)} K^{\pm}\pi^{\mp}$} and \mbox{$B^{0,\pm}\to \Dstar \hpm \pi^{\mp,0}$} contributions. The yields in individual bins are proportional to the $F_i$ parameters with the total yields for each contribution determined in the fit. 

The distributions of the misidentified contributions across the DP plane are the same as for the correctly identified ones. The total yields of these contributions are related to the corresponding correctly identified components using misidentification rates and selection efficiencies determined using calibration samples. Similarly, the yields of the incorrectly reconstructed $\Dstar$ contributions in both $D\piz$ and $D\gamma$ modes and partially reconstructed $\Dstar$ contributions in the $D\gamma$ modes are related to the correctly reconstructed cases by a freely varied ratio for each decay category that is assumed to be the same in individual DP bins and $B^{\pm}$ modes. These contributions also contribute to the $\gamma$ angle measurement. 

The total yields from correctly reconstructed \mbox{$B^{\pm}\rightarrow \Dstar \hpm, \Dstar\rightarrow \D\piz/\gamma, \D\rightarrow \KS \pi^+\pi^-$} and \mbox{$B^{\pm}\rightarrow \Dstar\hpm, \Dstar\rightarrow \D\piz/\gamma, \D\rightarrow \KS K^+K^-$} signal decays, obtained from the simultaneous 2D invariant-mass fit are shown in Tables~\ref{tab:yields_kspipi} and~\ref{tab:yields_kskk}, respectively. The total yields for the incorrectly and partially reconstructed $\Dstar$ contributions are provided in \cref{tab:indstyields_kspipi} and \cref{tab:indstyields_kskk} in \cref{app:Dstyields}. 

\begin{table}[]
    \centering
    \caption{Yields of fully reconstructed \Dstar contributions obtained from datasets with $\D\to \KS \pi^+\pi^-$ decays.}
    \begin{tabular}{c|c}
    Component & Yield \\\hline
$B^+ \rightarrow \Dstar \pi^+,\Dstar \rightarrow \D \piz$                 &$1273 \pm 32\phantom{1}$\\
$B^+ \rightarrow \Dstar \pi^+,\Dstar \rightarrow \D \gamma$               &$\phantom{}3692 \pm 158$\\
$B^- \rightarrow \Dstar \pi^-,\Dstar \rightarrow \D \piz$                 &$1290 \pm 33\phantom{1}$\\
$B^- \rightarrow \Dstar \pi^-,\Dstar \rightarrow \D \gamma$               &$\phantom{}3683 \pm 160$\\
$B^+ \rightarrow \Dstar K^+,\Dstar \rightarrow \D \piz$                   &$\phantom{1}112 \pm 7\phantom{11}$\\
$B^+ \rightarrow \Dstar K^+,\Dstar \rightarrow \D \gamma$                 &$\phantom{1}358 \pm 33\phantom{1}$\\
$B^- \rightarrow \Dstar K^-,\Dstar \rightarrow \D \piz$                   &$\phantom{1}109 \pm 6\phantom{11}$\\
$B^- \rightarrow \Dstar K^-,\Dstar \rightarrow \D \gamma$                 &$\phantom{1}419 \pm 35\phantom{1}$\\
\hline
    \end{tabular}
    \label{tab:yields_kspipi}
\end{table}

\begin{table}[]
    \centering
    \caption{Yields of fully reconstructed \Dstar contributions obtained from the datasets with \mbox{$\D\to \KS K^+K^-$} decays.}
    \begin{tabular}{c|c}
    Component & Yield \\ \hline
$B^+ \rightarrow \Dstar \pi^+,\Dstar \rightarrow \D \piz$                   &$\phantom{1}199 \pm 13\phantom{1}$\\
$B^+ \rightarrow \Dstar \pi^+,\Dstar \rightarrow \D \gamma$                  &$\phantom{1}782 \pm 49\phantom{1}$\\
$B^- \rightarrow \Dstar \pi^-,\Dstar \rightarrow \D \piz$                   &$\phantom{1}197 \pm 13\phantom{1}$\\
$B^- \rightarrow \Dstar \pi^-,\Dstar \rightarrow \D \gamma$                  &$\phantom{1}740 \pm 48\phantom{1}$\\
$B^+ \rightarrow \Dstar K^+,\Dstar \rightarrow \D \piz$                     &$\phantom{11}13 \pm 2\phantom{11}$\\
$B^+ \rightarrow \Dstar K^+,\Dstar \rightarrow \D \gamma$                    &$\phantom{11}69 \pm 11\phantom{1}$\\
$B^- \rightarrow \Dstar K^-,\Dstar \rightarrow \D \piz$                     &$\phantom{11}13 \pm 2\phantom{11}$\\
$B^- \rightarrow \Dstar K^-,\Dstar \rightarrow \D \gamma$                    &$\phantom{11}57 \pm 11\phantom{1}$\\
\hline
    \end{tabular}
    \label{tab:yields_kskk}
\end{table}

\begin{figure}
    \centering
    \includegraphics[width=0.47\textwidth]{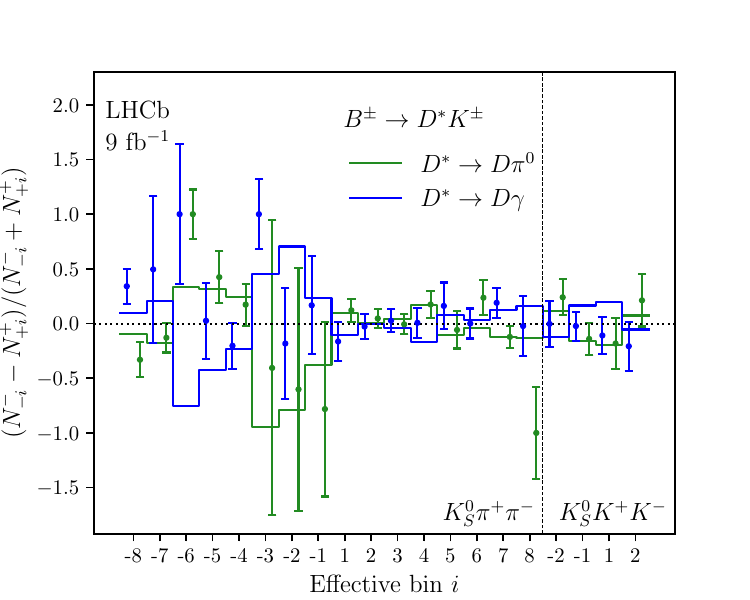}
    \includegraphics[width=0.47\textwidth]{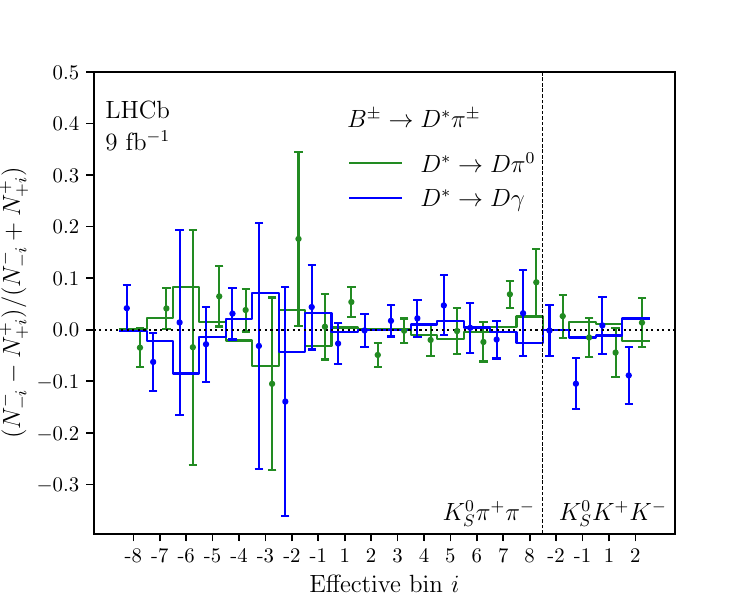}
    \caption{Expected asymmetry in each bin for (left) $B^{\pm} \to D^*K^{\pm}$ and (right) $B^{\pm} \to D^* \pi^{\pm}$, calculated from the \CP-violating observables measured in the default fit, are shown with the solid lines and obtained in fits with independent bin yields freely varied are shown with the error bars (data points). The vertical dashed lines separate the $\D \to \KS \pi^{+} \pi^{-}$ and $\D \to \KS K^{+} K^-$ bins on the horizontal axes.}
    \label{fig:CPV1}
\end{figure}

Alternative fits are performed to the mass distributions ($\mdh$ and $\mdn$) with the independent signal yields varied freely in individual DP bins. The yields are used to calculate the \CP asymmetries $(N^-_{-i} - N^+_{+i})/(N^-_{-i} + N^+_{+i})$ for effective bin pairs, defined to comprise bin $i$ for $B^+$ and bin $-i$ for $B^-$ decays. The \CP asymmetries are also calculated based on the \CP-violating observables determined in the default fit using \cref{eq:binned_yields}.  Figure~\ref{fig:CPV1} shows the comparison of the two results for the $B^{\pm}\rightarrow \Dstar K^{\pm}$ mode and the $B^{\pm}\rightarrow \Dstar \pi^{\pm}$ mode. 
No significant \CP-violation is observed in individual DP bins for $B^{\pm}\rightarrow \Dstar K^{\pm}$ due to limited statistics. 
The expected asymmetries are smaller in $B^{\pm}\rightarrow \Dstar \pi^{\pm}$. 

%% file: systematic.tex
\section{Systematic uncertainties}
\label{sec:systematic}

The systematic uncertainties are listed in \cref{tab:summary_systematic} and summarised below. In many cases, they are estimated using pseudoexperiments generated with inputs from simulations of the signal decay and employing amplitude model predictions~\cite{BaBar:2018cka}. 

The $m(\Dstar)$ and $m(Dh^{\pm})$ distributions are selected as the mass fit variables since they are much less correlated than \mdn and $m(\Dstar h)$. A small remnant correlation exists in the case of the partially reconstructed $B^{\pm}\rightarrow \Dstar h^{\pm}, \Dstar \rightarrow D\pi^0$ decay, in which one of the photons of the $\piz$ decay is missed. The effect is estimated by performing the fit with a model including the correlation and the biases in the \CP-violating observables are assigned as the systematic uncertainties. 

The presence of a nonuniform efficiency profile over the DP bins can change the average strong phase differences within a bin, \ie $c_i, s_i$. The size of the shifts of the $c_i, s_i$ values are estimated using efficiency profiles obtained from simulation and an amplitude model. Pseudoexperiments are performed where the events are generated using these shifted values and refitted with the nominal values. 

The uncertainties due to fixing the PDF parameters using fits to simulated samples with limited sample sizes are estimated through the resampling method~\cite{efron:1979}. 
Specifically, the datasets used to fix parameters are resampled and refit to obtain a new set of parameters. The fits to the data are then repeated. This process is performed multiple times, and the standard deviations of the fitted \CP-violating observables are assigned as systematic uncertainties. A similar method is used for the estimation of systematic uncertainties from other fixed parameters such as the branching ratios~\cite{PDG2022} and selection efficiencies used to constrain yields from different components.

The invariant-mass shapes used in the fits are assumed to be the same in all of the DP bins. To verify the validity of this assumption, the signal shapes are fit independently in each bin and only small differences are observed. The effect of ignoring these small differences is investigated using pseudoexperiments generated according to the bin-dependent PDF parameters. The data is subsequently fit with the default model and the average biases for each \CP-violating observable are assigned as systematic uncertainties.

Resolution effects in the reconstructed kinematic variables can lead to bin migration among DP bins. Most of this effect is accounted for in the nominal fits, since the $F_i$ parameters are determined directly from the data. However, there may be residual effects due to differences in the $\Bpm\to\Dstar \Kpm$ and $\Bpm\to\Dstar \pipm$ distributions. The effect is estimated by generating pseudoexperiments in which the DP coordinates are smeared using the experimental resolution determined by simulation. The DP bins are labelled based on these smeared coordinates in the DP using the binning scheme shown in \cref{fig:binning} and the fits are run ignoring this effect. The mean biases of the extracted \CP-violating observables are assigned as systematic uncertainties.

To estimate the related systematic uncertainty due to $\Lb$ decay backgrounds,  pseudoexperiments are generated with $\Lb$ background contributions and then refitted with the default model. The mean biases of the \CP-violating observables are assigned as systematic uncertainties. The systematic uncertainty due to the crossfeed from $\Lb\rightarrow D^0p\pi$ is one of the largest contributions. 
The systematic uncertainties from semileptonic decays are also estimated similarly and the effect is negligible compared to the statistical uncertainties.

Some data subsamples across the \KS reconstructed categories are merged in the default fit because of the small difference between various data subsamples. The bias caused by the merged data subsamples is estimated by the alternative simultaneous fit to the various data subsamples. The bias between the default fit and alternative simultaneous fit is assigned as systematic uncertainties. The systematic uncertainties caused by the merged data subsamples is negligible compared to the statistical uncertainties.

Pseudoexperiments are performed in which the datasets are generated according to the world average values for $\ckmgammaangle$ and the hadronic parameters~\cite{HFLAV:2022pwe}. No bias is observed in the fitted values and thus no additional systematic uncertainty is assigned. 
In the default fit, \CP-violation is not considered for the $B^{\pm,0}\rightarrow DK^{\pm}\pi^{0,\mp}$ contributions; the effect of this assumption is estimated by running the fit with the model allowing for \CP-violation in this decay. The difference is assigned as a systematic uncertainty.

In the fit, the parameters $c_i$ and $s_i$ are fixed to the values measured by the \besiii and \cleo collaborations~\cite{Ablikim:2020lpk, Ablikim:2020cfp, PhysRevD.82.112006}. The systematic uncertainties associated with the precision of these external measurements are estimated by performing multiple fits to the data with different sets of $c_i$ and $s_i$ values, generated according to their uncertainties and correlations. The standard deviations of the distributions of the fitted \CP-violating observables are assigned as the corresponding systematic uncertainties. 

In summary, the systematic uncertainties from various sources are small compared to the statistical uncertainties. The efficiency effects, effects from $\Lb$ decays as well as the fixed invariant-mass shapes used in all DP bins are the dominant sources of systematic uncertainties. 
\begin{table}[]
    \centering
    \caption{Summary of uncertainties on the measurement of $x^{\Dstar K}_{\pm},y^{\Dstar K}_{\pm},x^{\Dstar \pi}_{\xi}$ and $y^{\Dstar \pi}_{\xi}$. All numbers have been scaled up by a factor of 100.}
    \resizebox{\textwidth}{!}{
    \begin{tabular}{c|c|c|c|c|c|c}
    \multicolumn{7}{c}{} \\
    Source                                                              &  $\sigma(x^{\Dstar K}_+)$ & $\sigma(x^{\Dstar K}_-)$ & $\sigma(y^{\Dstar K}_+)$ & $\sigma(y^{\Dstar K}_-)$ & $\sigma(x_{\xi}^{\Dstar\pi})$ & $\sigma(y_{\xi}^{\Dstar\pi})$ \\ \hline
 
    Neglecting correlations                                             & 0.05                      & 0.03                     & 0.19                     & 0.04                     & 0.70                          & 1.48 \\ 
    Efficiency correction of ($c_i,s_i$)                                & 0.53                      & 0.18                      & 0.18                    & 0.20                     & 0.64                          & 1.73 \\ 
    Invariant mass shape parameter                                      & 0.09                      & 0.16                      & 0.20                    & 0.05                     & 0.39                          & 0.06 \\
    Fixed yield ratios                                                  & 0.09                      & 0.03                      & 0.03                    & 0.01                     & 0.33                          & 0.15\\
    Bin dependence of the invariant-mass shape                          & 0.40                      & 0.38                      & 0.41                    & 0.33                     & 1.78                          & 1.57 \\
    DP bin migration                                                    & 0.32                      & 0.70                      & 0.03                    & 0.17                     & 1.2                           & 2.0\\
    $\Lb$ background                                                    & 0.97                      & 1.34                      & 0.55                    & 0.77                     & 1.13                          & 1.43\\
    Semileptonic $B$ backgrounds                                        & 0.27                      & 1.29                      & 0.02                    & 0.67                     & 0.03                          & 0.04 \\
    Merging data subsamples                                             & 0.06                      & 0.02                      & 0.12                    & 0.03                     & 0.06                          & 0.34 \\
    \CP-violation in $B^{\pm,0}\rightarrow DK^{\pm}\pi^{0,\mp}$         & 0.03                      & 0.13                      & 1.97                    & 0.99                     & 0.13                          & 0.68 \\
    \hline
    Total systematic                                                    & \xpvsys                   & \xmvsys                   & \ypvsys         
           & \ymvsys                  & \xxivsys                      & \yxivsys \\
    \hline
    Strong-phase inputs (external)                                      & 0.41                      & 0.23                      & 0.30                    & 0.64                     & 0.93                          & 0.83 \\
    \hline
    Statistical                                                         & \xpe                      & \xme                      & \ype                    & \yme                     & \xxie                         & \yxie \\ \hline
    \end{tabular}
    }
    \label{tab:summary_systematic}
\end{table}

%% file: interpret.tex
\section{Interpretation}
\label{sec:interpretation}
\begin{figure}
    \centering
    \includegraphics[width=0.48\textwidth]{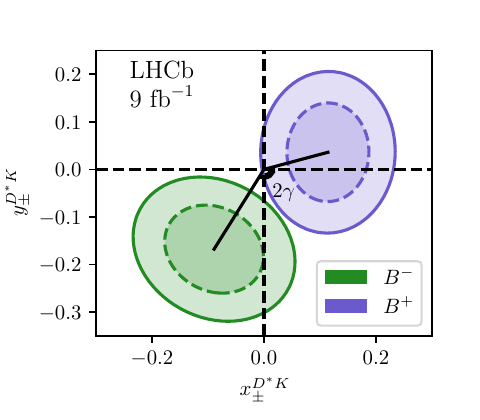}
    \includegraphics[width=0.48\textwidth]{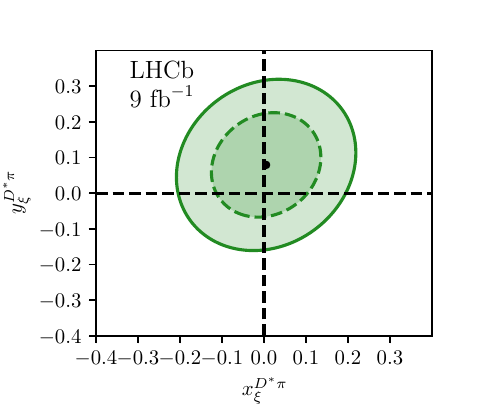}
    \caption{Confidence level at 68.2\% and 95.5\% probability for (left, green) $x^{\Dstar K}_{+},y^{\Dstar K}_{+}$, (left, violet) $x^{\Dstar K}_{-},y^{\Dstar K}_{-}$ and (right, green) $x^{\Dstar \pi}_{\xi},y^{\Dstar \pi}_{\xi}$ are measured in $B^{\pm}\to \Dstar h^{\pm}$ decays from a profile likelihood scan.}
    \label{fig:LLscan}
\end{figure}

The measured \CP-violating observables are 
\begin{equation*}
    \begin{aligned}
    \label{eq:xyres}
        x^{\Dstar K}_+ &=& (\xpvstat \pm \xpvsys \pm 0.41) \times 10^{-2},  \\
        x^{\Dstar K}_- &=& (\xmvstat \pm \xmvsys \pm 0.23) \times 10^{-2},  \\
        y^{\Dstar K}_+ &=& (\ypvstat \pm \ypvsys \pm 0.30) \times 10^{-2},  \\
        y^{\Dstar K}_- &=& (\ymvstat \pm \ymvsys \pm 0.64) \times 10^{-2}, \\
        x_{\xi}^{\Dstar \pi} &=& (\xxivstat \pm \xxivsys \pm 0.93) \times 10^{-2}, \\
        y_{\xi}^{\Dstar \pi} &=& (\yxivstat \pm \yxivsys \pm 0.83) \times 10^{-2}, \\ 
    \end{aligned}
\end{equation*}
where the first uncertainty is statistical, the second is the systematic uncertainty from \lhcb-related sources including the analysis method and detector effects, and the third is from the external strong phase inputs from the \besiii and \cleo collaborations~\cite{Ablikim:2020lpk, Ablikim:2020cfp, PhysRevD.82.112006}. The correlation matrices for all three uncertainties are provided in \cref{cormatrix}.
The two-dimensional profile likelihood scans are shown on the left and right in \cref{fig:LLscan} for ($x^{\Dstar K}_{\pm},y^{\Dstar K}_{\pm}$) and ($x^{\Dstar \pi}_{\xi},y^{\Dstar \pi}_{\xi}$), respectively.

\begin{figure}
    \centering\includegraphics[width=0.55\textwidth]{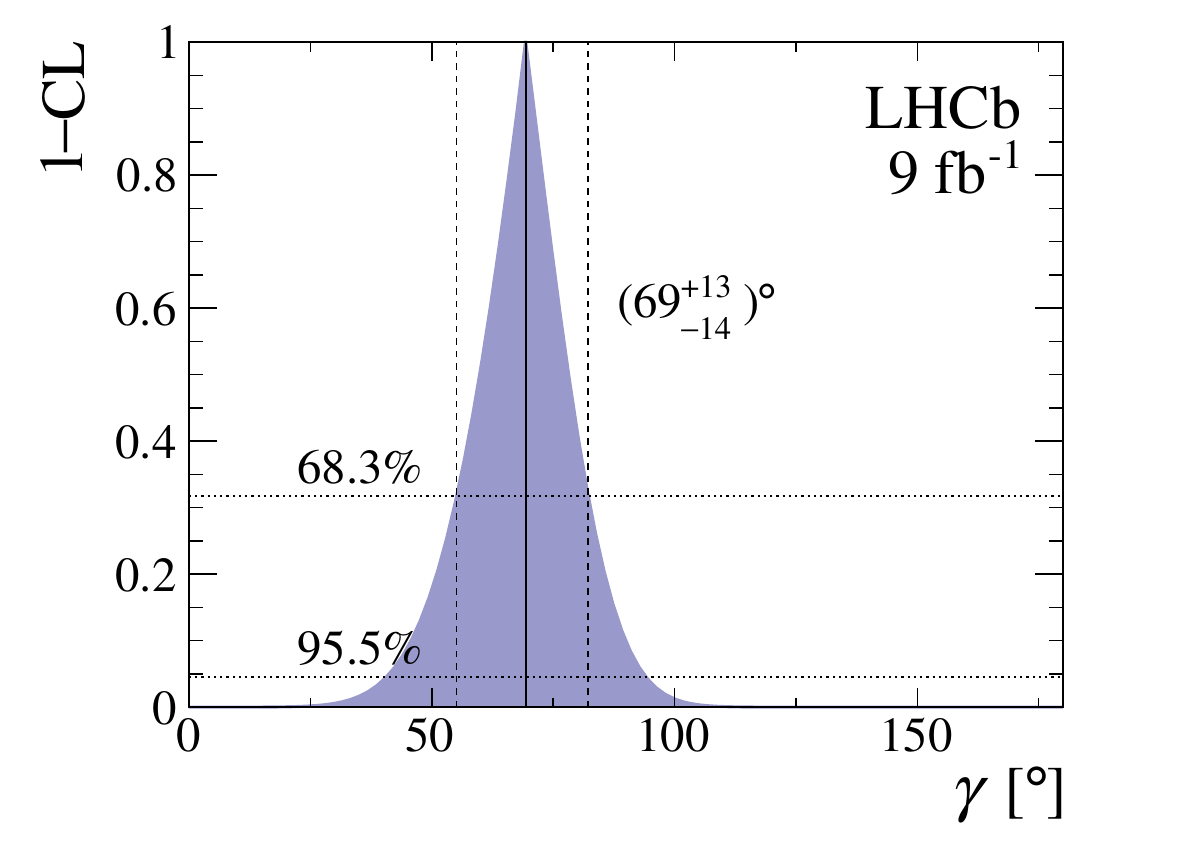}
    \caption{Confidence limits for the CKM angle $\gamma$, as determined from {\tt GammaCombo}~\cite{GammaCombo, LHCb-PAPER-2021-033}.}
    \label{fig:gamma_angle}
\end{figure}

\begin{figure}
    \centering
    \includegraphics[width=0.48\textwidth]{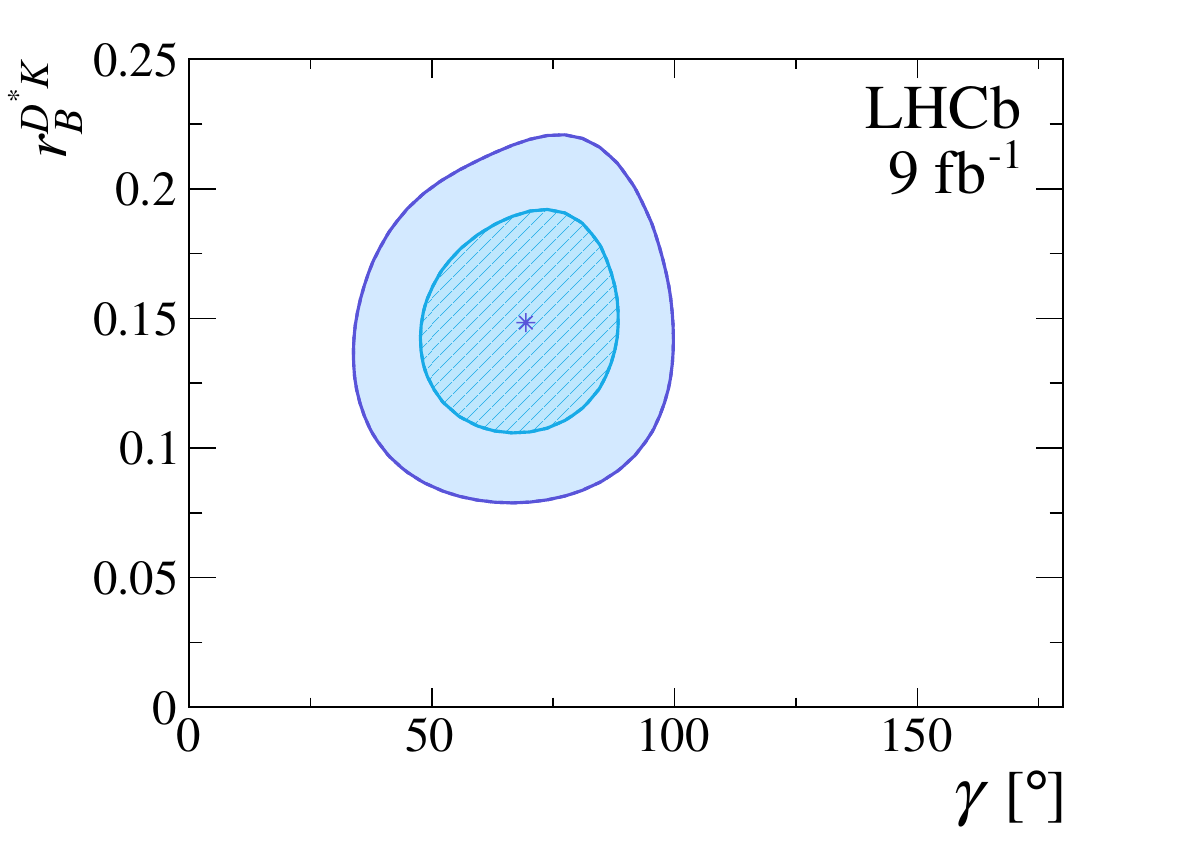}
    \includegraphics[width=0.48\textwidth]{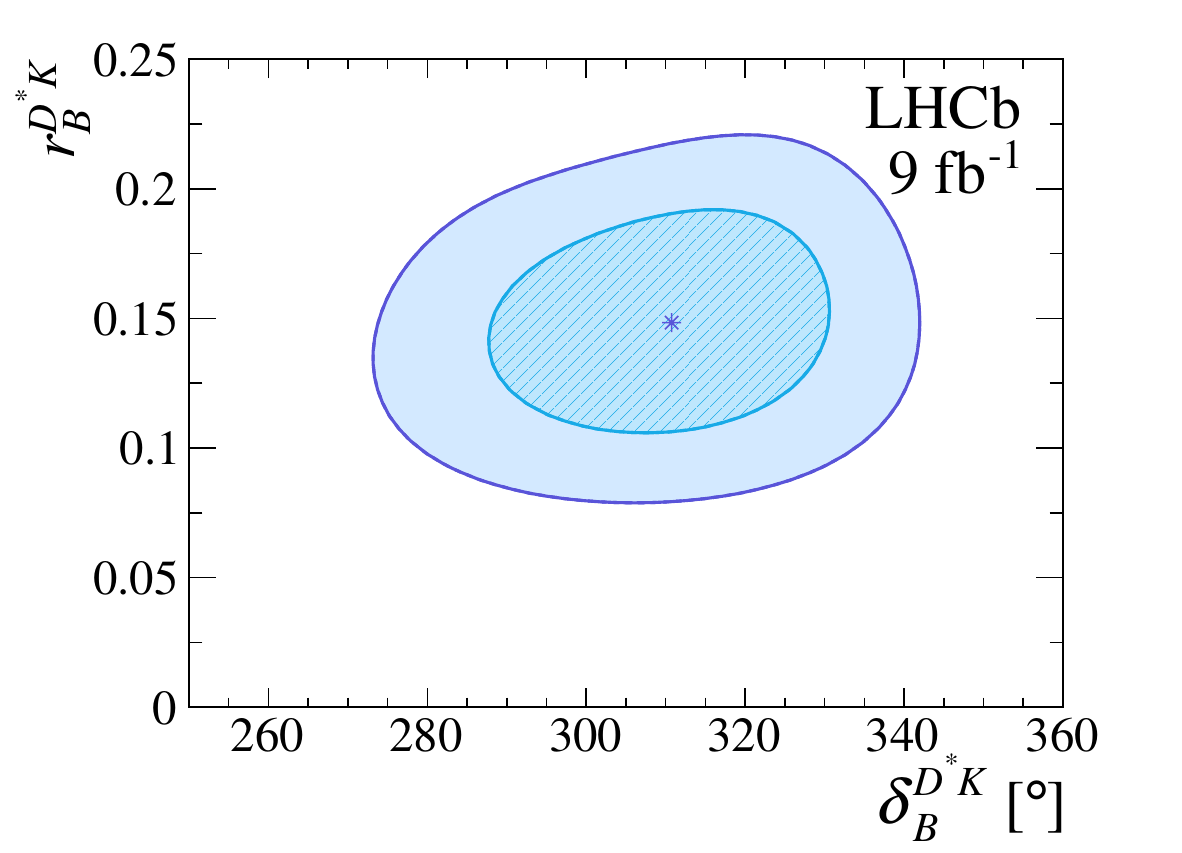}
    \includegraphics[width=0.48\textwidth]{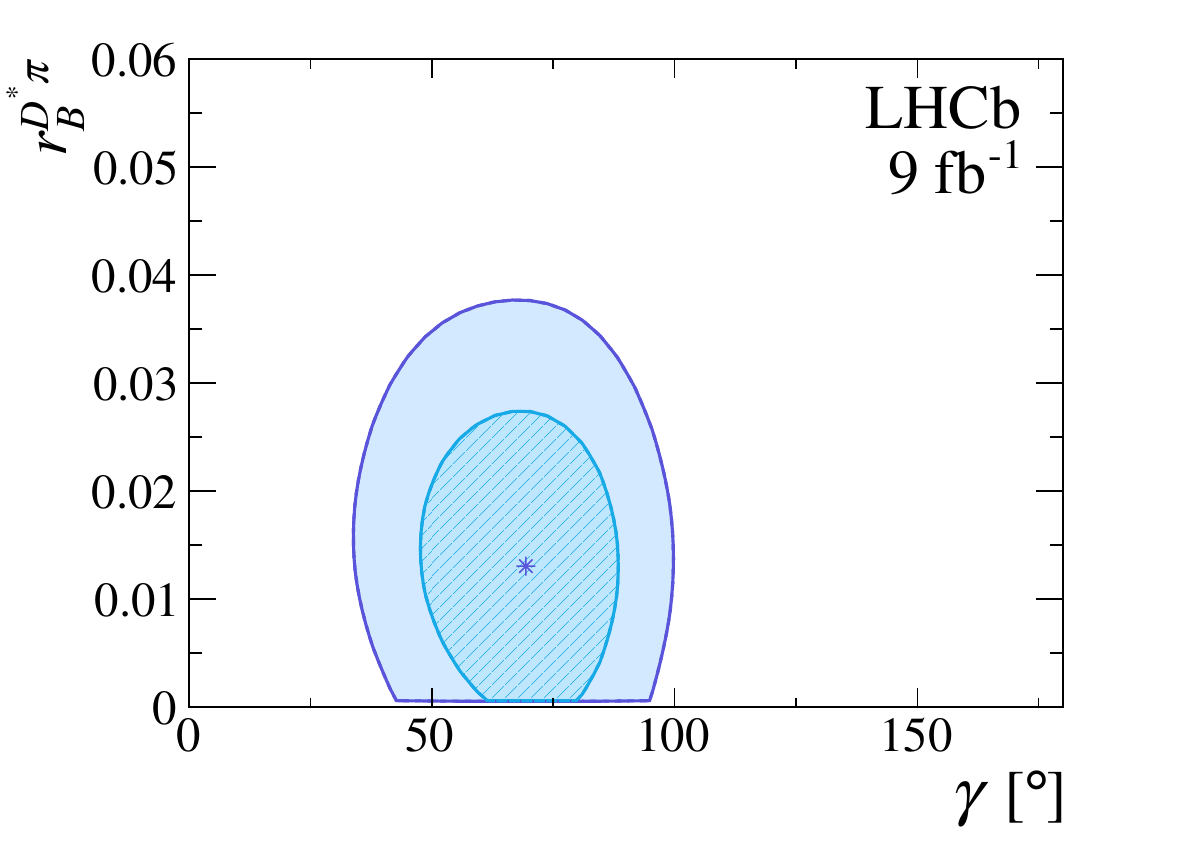}
    \includegraphics[width=0.48\textwidth]{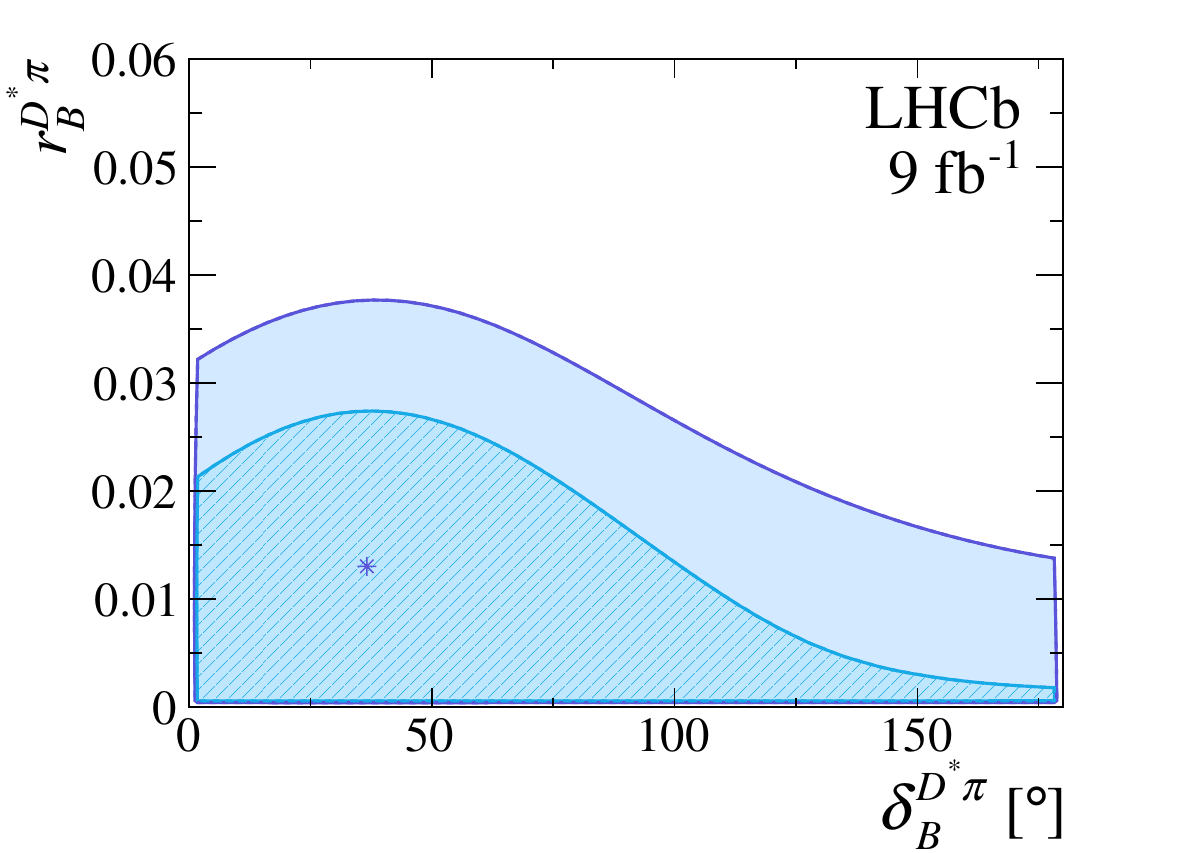}
    \caption{Confidence level regions for the combination of physical parameters ($\ckmgammaangle, \delta_B^{\Dstar K}, r_B^{\Dstar K}, \delta_B^{\Dstar \pi}, r_B^{\Dstar \pi}$) of interest. The 68\% (hashed blue area) and 95\% (light blue area) are determined from {\tt GammaCombo}~\cite{GammaCombo, LHCb-PAPER-2021-033}. }
    \label{fig:2D_interpretation}
\end{figure}

The measured values of $x^{\Dstar K}_{\pm},y^{\Dstar K}_{\pm}$, $x^{\Dstar \pi}_{\xi}$ and $y^{\Dstar \pi}_{\xi}$ are interpreted in terms of the parameters of interest: $\ckmgammaangle, r_{B}^{\Dstar  K}, r_B^{\Dstar \pi}, \delta_B^{\Dstar  K}$ and $\delta_B^{\Dstar  \pi}$. The interpretation and confidence intervals are calculated via an extended maximum likelihood fit using a Bayesian approach while determining $\gamma$ with the PROB method used in the {\tt GammaCombo} package~\cite{GammaCombo, matthew_kenzie_2021_5503651, 
LHCb-PAPER-2021-033}. The obtained results are:
\begin{equation*}
    \begin{aligned}
     \gamma &= \vargamma, \\
     r_B^{\Dstar K} &= \varrbdstk, \\
     r_B^{\Dstar \pi} &= \varrbdstpi, \\
     \delta_B^{\Dstar K} &= \vardeltabdstk,   \\
     \delta_B^{\Dstar \pi} &= \vardeltabdstpi.  \\ 
    \end{aligned}
\end{equation*}
The combined uncertainty in the $\ckmgammaangle$ measurement is dominated by the statistical uncertainties and is larger than that in the $B^{\pm}\to Dh^{\pm}$ measurement~\cite{LHCb-PAPER-2020-019}. 
The one-dimensional confidence limits plots are shown in \cref{fig:gamma_angle}, and the two-dimensional plots showing the confidence level regions are shown in \cref{fig:2D_interpretation}. 

\clearpage

%% file: conclude.tex
\section{Conclusion}
\label{sec:conclude}
In summary, the $\ckmanglegamma$ is measured using a data sample of \mbox{ $\Bpm\rightarrow \Dstar \hpm, \Dstar \rightarrow \D\piz/\gamma, \D \rightarrow \KS h^+h^- (h=K,\pi)$} decays collected from $pp$ collisions at the \lhcb experiment corresponding to an integrated luminosity of 9 \invfb at centre-of-mass energies of 7, 8 and 13 \tev. The analysis is performed in bins of the $\D \to\KS\hp\hm$ DP variables, utilising the strong phase information measured by the \besiii and \cleo collaborations~\cite{Ablikim:2020lpk, Ablikim:2020cfp, PhysRevD.82.112006}. Interpretation in terms of $\gamma$ and hadronic parameters of the $B^{\pm}$ decays yields $\gamma = \vargamma$. These results are consistent with the world average~\cite{HFLAV:2022pwe} and those from the $B^{\pm}\rightarrow D^{(*)}h^{\pm}$ measurements~\cite{LHCb-PAPER-2020-019, LHCb-PAPER-2020-036, Poluektov:2004mf}. The measurement gives the most precise determination to date of $\ckmgammaangle$ using these channels and will help improve the sensitivity in the global fit to $\gamma$.

\clearpage

%% file: acknowledgements.tex
\section*{Acknowledgements}

\noindent We express our gratitude to our colleagues in the CERN
accelerator departments for the excellent performance of the LHC. We
thank the technical and administrative staff at the LHCb
institutes.
We acknowledge support from CERN and from the national agencies:
CAPES, CNPq, FAPERJ and FINEP (Brazil); 
MOST and NSFC (China); 
CNRS/IN2P3 (France); 
BMBF, DFG and MPG (Germany); 
INFN (Italy); 
NWO (Netherlands); 
MNiSW and NCN (Poland); 
MCID/IFA (Romania); 
MICINN (Spain); 
SNSF and SER (Switzerland); 
NASU (Ukraine); 
STFC (United Kingdom); 
DOE NP and NSF (USA).
We acknowledge the computing resources that are provided by CERN, IN2P3
(France), KIT and DESY (Germany), INFN (Italy), SURF (Netherlands),
PIC (Spain), GridPP (United Kingdom), 
CSCS (Switzerland), IFIN-HH (Romania), CBPF (Brazil),
Polish WLCG  (Poland) and NERSC (USA).
We are indebted to the communities behind the multiple open-source
software packages on which we depend.
Individual groups or members have received support from
ARC and ARDC (Australia);
Minciencias (Colombia);
AvH Foundation (Germany);
EPLANET, Marie Sk\l{}odowska-Curie Actions, ERC and NextGenerationEU (European Union);
A*MIDEX, ANR, IPhU and Labex P2IO, and R\'{e}gion Auvergne-Rh\^{o}ne-Alpes (France);
Key Research Program of Frontier Sciences of CAS, CAS PIFI, CAS CCEPP, 
Fundamental Research Funds for the Central Universities, 
and Sci. \& Tech. Program of Guangzhou (China);
GVA, XuntaGal, GENCAT, Inditex, InTalent and Prog.~Atracci\'on Talento, CM (Spain);
SRC (Sweden);
the Leverhulme Trust, the Royal Society
 and UKRI (United Kingdom).

%% file: appendix.tex
\section*{Appendices}

\appendix
\input{cor_matrix.tex}

\input{yields}

%% file: cor_matrix.tex
\section{Correlation matrices}
\label{cormatrix}

The statistical correlation matrix can be found in \cref{tab:stat_cor_mat}. The systematic correlation matrices for the \lhcb related effects and the strong phase inputs are shown in \cref{tab:sys_cor_mat} and \cref{tab:sys_cor_mat_strph}, respectively.

\begin{table}[hb]
    \centering
    \caption{Statistical correlation matrix for \CP-violating observables.}
    \begin{tabular}{c|cccccc}
                              &  $x^{D^*\pi}_{\xi}$  &  $x^{D^*K}_-$  & $x^{D^*K}_+$   & $y^{D^*\pi}_{\xi}$      & $y^{D^*K}_-$     & $y^{D^*K}_+$ \\
                              \hline
    $x^{D^*\pi}_{\xi}$        &    1                 &  0.25          &  $-0.16$       &  0.18                   &  $0.00$          & $0.00$ \\

    $x^{D^*K}_-$              &                      &  1             &  $-0.05$       &  0.06                   &  $-0.08$         & 0.03   \\

    $x^{D^*K}_+$              &                      &                & 1              & $-0.14$                 & $-0.08$          & $-0.08$\\

    $y^{D^*\pi}_{\xi}$        &                      &                &                &   1                     &  0.33            & $-0.19$\\

    $y^{D^*K}_-$              &                      &                &                &                         &   1              & $-0.09$\\

    $y^{D^*K}_+$              &                      &                &                &                         &                  &  1   \\
    \end{tabular}
    \label{tab:stat_cor_mat}
\end{table}

\begin{table}[hb]
    \centering
    \caption{Systematic correlation matrix for considered \lhcb related effects.}
    \begin{tabular}{c|cccccc}
                        &  $x^{D^*K}_-$  &  $x^{D^*K}_+$  & $y^{D^*K}_-$ & $y^{D^*K}_+$ & $x^{D^*\pi}_{\xi}$ & $y^{D^*\pi}_{\xi}$ \\
                    \hline
    $x^{D^*K}_-$        &    1           &  $-0.01$       & 0.01         & 0.02         &  $-0.02$           & $-0.05$ \\
    $x^{D^*K}_+$        &                &  1             & $-0.01$      & 0.00         & $-0.01$            & 0.06 \\
    $y^{D^*K}_-$        &                &                & 1            & $-0.01$      & 0.01               & $-0.10$\\
    $y^{D^*K}_+$        &                &                &              &   1          & $-0.09$            & $-0.02$\\
    $x^{D^*\pi}_{\xi}$  &                &                &              &              &   1                & 0.02\\
    $y^{D^*\pi}_{\xi}$  &                &                &              &              &                    &  1   \\
    \end{tabular}
    \label{tab:sys_cor_mat}
\end{table}

\begin{table}[hb]
    \centering
    \caption{Systematic correlation matrix for the strong phase inputs.}
    \begin{tabular}{c|cccccc}
                        &  $x^{D^*K}_-$  &  $x^{D^*K}_+$  & $y^{D^*K}_-$ & $y^{D^*K}_+$ & $x^{D^*\pi}_{\xi}$ & $y^{D^*\pi}_{\xi}$ \\
                     \hline
    $x^{D^*K}_-$        &    1           &  $0.26 $       & $-0.58$      & $-0.21$      &  $-0.53$           & $-0.39$ \\
    $x^{D^*K}_+$        &                &  1             & $0.07$       & $0.44$       &  $-0.59$           & $0.24$ \\
    $y^{D^*K}_-$        &                &                & 1            & $-0.04$      & $-0.24 $           & $-0.03$\\
    $y^{D^*K}_+$        &                &                &              &   1          & $0.08$             & $0.44$\\
    $x^{D^*\pi}_{\xi}$  &                &                &              &              &   1                & $0.57$\\
    $y^{D^*\pi}_{\xi}$  &                &                &              &              &                    &  1   \\
    \end{tabular}
    \label{tab:sys_cor_mat_strph}
\end{table}

\clearpage

%% file: yields.tex
\section{Incorrectly reconstructed \texorpdfstring{\Dstar}{Dstar} yields }
\label{app:Dstyields}

The incorrectly reconstructed $\Dstar$ contributions in $\piz/\gamma$ modes and partially reconstructed $\Dstar$ contributions in $\gamma$ modes related to the signal decays contribute sensitivity to the CKM angle $\gamma$ measurement. The yields of the contributions are shown in \cref{tab:indstyields_kspipi} and \cref{tab:indstyields_kskk} for the $\D \to \KS \pi^+ \pi^-$ and $\D \to \KS K^+ K^-$ categories, respectively. 

\begin{table}[]
    \centering
    \caption{Yields of incorrectly reconstructed and partially reconstructed \Dstar contributions from datasets with $D\to \KS \pi^+\pi^-$ decays.}
    \begin{tabular}{c|c}
    Component                                                                    & Yields \\\hline
$B^+ \rightarrow (\D\piz)_{\Dstar} \pi^+, \D + \textrm{random } \piz$                          &$\phantom{1}637 \pm 41 \phantom{1}$ \\
$B^+ \rightarrow (\D\gamma)_{\Dstar} \pi^+, \D + \textrm{random } \piz$                        &$295 \pm 39$ \\
$B^+ \rightarrow (\D\piz)_{\Dstar} \pi^+, \D + \textrm{random } \gamma$                          &$2877 \pm 146$ \\
$B^+ \rightarrow (\D\gamma)_{\Dstar} \pi^+, \D + \textrm{random } \gamma$                    &$2880 \pm 253$ \\
$B^+ \rightarrow (\D\piz)_{\Dstar} \pi^+, \D + \textrm{correct } \gamma$                    &$4417 \pm 179$ \\

$B^- \rightarrow (\D\piz)_{\Dstar} \pi^-, \D + \textrm{random } \piz$              &$\phantom{1}645 \pm 42\phantom{1}$ \\
$B^- \rightarrow (\D\gamma)_{\Dstar} \pi^-, \D + \textrm{random } \piz$             &$295 \pm 39$ \\
$B^- \rightarrow (\D\piz)_{\Dstar} \pi^-, \D + \textrm{random } \gamma$               &$2915 \pm 149$ \\
$B^- \rightarrow (\D\gamma)_{\Dstar} \pi^-, \D + \textrm{random } \gamma$             &$2873 \pm 254$ \\
$B^- \rightarrow (\D\piz)_{\Dstar} \pi^-, \D + \textrm{correct } \gamma$             &$4476 \pm 182$ \\

$B^+ \rightarrow (\D\piz)_{\Dstar} K^+, \D + \textrm{random } \piz$                 &$\phantom{1}56 \pm 5\phantom{1}$ \\
$B^+ \rightarrow (\D\gamma)_{\Dstar} K^+, \D + \textrm{random } \piz$               &$ 29 \pm 4$ \\
$B^+ \rightarrow (\D\piz)_{\Dstar} K^+, \D + \textrm{random } \gamma$                &$ 253 \pm 19$ \\
$B^+ \rightarrow (\D\gamma)_{\Dstar} K^+, \D + \textrm{random } \gamma$               &$ 279 \pm 34$ \\
$B^+ \rightarrow (\D\piz)_{\Dstar} K^+, \D + \textrm{correct } \gamma$              &$ 389 \pm 27$ \\

$B^- \rightarrow (\D\piz)_{\Dstar} K^-, \D + \textrm{random } \piz$                 &$ 55 \pm 4$ \\
$B^- \rightarrow (\D\gamma)_{\Dstar} K^-, \D + \textrm{random } \piz$               &$ 34 \pm 5$ \\
$B^- \rightarrow (\D\piz)_{\Dstar} K^-, \D + \textrm{random } \gamma$                 &$ 246 \pm 17$ \\
$B^- \rightarrow (\D\gamma)_{\Dstar} K^-, \D + \textrm{random } \gamma$               &$ 327 \pm 37$ \\
$B^- \rightarrow (\D\piz)_{\Dstar} K^-, \D + \textrm{correct } \gamma$               &$ 378 \pm 24$ \\
    \end{tabular}
    \label{tab:indstyields_kspipi}
\end{table}

\begin{table}[]
    \centering
    \caption{Yields of incorrectly reconstructed and partially reconstructed \Dstar contributions from the datasets with $\D\to K_S^0K^+K^-$ decays.}
    \begin{tabular}{c|c}
    Component                                                          & Yields \\ \hline
$B^+ \rightarrow (\D\piz)_{\Dstar} \pi^+, \D + \textrm{random } \piz$            &$ 103 \pm 15$ \\
$B^+ \rightarrow (\D\gamma)_{\Dstar} \pi^+, \D + \textrm{random } \piz$          &$\phantom{1}86 \pm  17$ \\
$B^+ \rightarrow (\D\piz)_{\Dstar} \pi^+, \D + \textrm{random } \gamma$            &$ 330 \pm  35$ \\
$B^+ \rightarrow (\D\gamma)_{\Dstar} \pi^+, \D + \textrm{random } \gamma$          &$ 500 \pm  92$ \\
$B^+ \rightarrow (\D\piz)_{\Dstar} \pi^+, \D + \textrm{correct } \gamma$          &$ 923 \pm  91$ \\

$B^- \rightarrow (\D\piz)_{\Dstar} \pi^-, \D + \textrm{random } \piz$    &$ 102 \pm 15$ \\
$B^- \rightarrow (\D\gamma)_{\Dstar} \pi^-, \D + \textrm{random } \piz$   &$\phantom{1}81 \pm 16$ \\
$B^- \rightarrow (\D\piz)_{\Dstar} \pi^-, \D + \textrm{random } \gamma$     &$ 327 \pm 35$ \\
$B^- \rightarrow (\D\gamma)_{\Dstar} \pi^-, \D + \textrm{random } \gamma$   &$ 474 \pm  87$ \\
$B^- \rightarrow (\D\piz)_{\Dstar} \pi^-, \D + \textrm{correct } \gamma$   &$ 914 \pm  90$ \\

$B^+ \rightarrow (\D\piz)_{\Dstar} K^+, \D + \textrm{random } \piz$       &$\phantom{1}7 \pm 1$ \\
$B^+ \rightarrow (\D\gamma)_{\Dstar} K^+, \D + \textrm{random } \piz$     &$\phantom{1}8 \pm 2$ \\
$B^+ \rightarrow (\D\piz)_{\Dstar} K^+, \D + \textrm{random } \gamma$      &$ 22 \pm  4$ \\
$B^+ \rightarrow (\D\gamma)_{\Dstar} K^+, \D + \textrm{random } \gamma$     &$\phantom{1}44 \pm 10$ \\
$B^+ \rightarrow (\D\piz)_{\Dstar} K^+, \D + \textrm{correct } \gamma$    &$\phantom{1}60 \pm 10$ \\

$B^- \rightarrow (\D\piz)_{\Dstar} K^-, \D + \textrm{random } \piz$       &$\phantom{1}7 \pm 1$ \\
$B^- \rightarrow (\D\gamma)_{\Dstar} K^-, \D + \textrm{random } \piz$     &$\phantom{1}6 \pm 2$ \\
$B^- \rightarrow (\D\piz)_{\Dstar} K^-, \D + \textrm{random } \gamma$       &$ 22 \pm 4$ \\
$B^- \rightarrow (\D\gamma)_{\Dstar} K^-, \D + \textrm{random } \gamma$     &$ 36 \pm 9$ \\
$B^- \rightarrow (\D\piz)_{\Dstar} K^-, \D + \textrm{correct } \gamma$     &$\phantom{1}60 \pm 10$ \\

    \end{tabular}
    \label{tab:indstyields_kskk}
\end{table}

\clearpage

%% file: Authorship_LHCb-PAPER-2023-012.tex
\centerline
{\large\bf LHCb collaboration}
\begin
{flushleft}
\small
R.~Aaij$^{32}$\lhcborcid{0000-0003-0533-1952},
A.S.W.~Abdelmotteleb$^{51}$\lhcborcid{0000-0001-7905-0542},
C.~Abellan~Beteta$^{45}$,
F.~Abudin{\'e}n$^{51}$\lhcborcid{0000-0002-6737-3528},
T.~Ackernley$^{55}$\lhcborcid{0000-0002-5951-3498},
B.~Adeva$^{41}$\lhcborcid{0000-0001-9756-3712},
M.~Adinolfi$^{49}$\lhcborcid{0000-0002-1326-1264},
P.~Adlarson$^{77}$\lhcborcid{0000-0001-6280-3851},
H.~Afsharnia$^{9}$,
C.~Agapopoulou$^{43}$\lhcborcid{0000-0002-2368-0147}
% ...

{\footnotesize \it

% ...
$^{9}$Universit{\'e} Clermont Auvergne, CNRS/IN2P3, LPC, Clermont-Ferrand, France\\
% ...
$^{32}$Nikhef National Institute for Subatomic Physics, Amsterdam, Netherlands\\
% ...
$^{41}$Instituto Galego de F{\'\i}sica de Altas Enerx{\'\i}as (IGFAE), Universidade de Santiago de Compostela, Santiago de Compostela, Spain\\
% ...
$^{43}$European Organization for Nuclear Research (CERN), Geneva, Switzerland\\
% ...
$^{45}$Physik-Institut, Universit{\"a}t Z{\"u}rich, Z{\"u}rich, Switzerland\\
% ...
$^{49}$H.H. Wills Physics Laboratory, University of Bristol, Bristol, United Kingdom\\
% ...
$^{51}$Department of Physics, University of Warwick, Coventry, United Kingdom\\
% ...
$^{54}$School of Physics and Astronomy, University of Glasgow, Glasgow, United Kingdom\\
$^{55}$Oliver Lodge Laboratory, University of Liverpool, Liverpool, United Kingdom\\
% ...
$^{77}$Department of Physics and Astronomy, Uppsala University, Uppsala, Sweden, associated to $^{54}$\\
% ...
}
\end{flushleft}

%% file: main.bbl
\ifx\mcitethebibliography\mciteundefinedmacro
\PackageError{LHCb.bst}{mciteplus.sty has not been loaded}
{This bibstyle requires the use of the mciteplus package.}\fi
\providecommand{\href}[2]{#2}
\begin{mcitethebibliography}{10}
\mciteSetBstSublistMode{n}
\mciteSetBstMaxWidthForm{subitem}{\alph{mcitesubitemcount})}
\mciteSetBstSublistLabelBeginEnd{\mcitemaxwidthsubitemform\space}
{\relax}{\relax}

\bibitem{PhysRevLett.10.531}
N.~Cabibbo, \ifthenelse{\boolean{articletitles}}{\emph{Unitary symmetry and leptonic decays}, }{}\href{https://doi.org/10.1103/PhysRevLett.10.531}{Phys.\ Rev.\ Lett.\  \textbf{10} (1963) 531}\relax
\mciteBstWouldAddEndPuncttrue
\mciteSetBstMidEndSepPunct{\mcitedefaultmidpunct}
{\mcitedefaultendpunct}{\mcitedefaultseppunct}\relax
\EndOfBibitem
\bibitem{10.1143/PTP.49.652}
M.~Kobayashi and T.~Maskawa, \ifthenelse{\boolean{articletitles}}{\emph{{CP-Violation in the renormalizable theory of weak interaction}}, }{}\href{https://doi.org/10.1143/PTP.49.652}{Progress of Theoretical Physics \textbf{49} (1973) 652}\relax
\mciteBstWouldAddEndPuncttrue
\mciteSetBstMidEndSepPunct{\mcitedefaultmidpunct}
{\mcitedefaultendpunct}{\mcitedefaultseppunct}\relax
\EndOfBibitem
\bibitem{Brod:2013sga}
J.~Brod and J.~Zupan, \ifthenelse{\boolean{articletitles}}{\emph{{The ultimate theoretical error on $\gamma$ from $B \to DK$ decays}}, }{}\href{https://doi.org/10.1007/JHEP01(2014)051}{JHEP \textbf{01} (2014) 051}, \href{http://arxiv.org/abs/1308.5663}{{\normalfont\ttfamily arXiv:1308.5663}}\relax
\mciteBstWouldAddEndPuncttrue
\mciteSetBstMidEndSepPunct{\mcitedefaultmidpunct}
{\mcitedefaultendpunct}{\mcitedefaultseppunct}\relax
\EndOfBibitem
\bibitem{Blanke:2018cya}
M.~Blanke and A.~J. Buras, \ifthenelse{\boolean{articletitles}}{\emph{{Emerging $\Delta M_{d}$ anomaly from tree-level determinations of $|V_{cb}|$ and the angle $\gamma $}}, }{}\href{https://doi.org/10.1140/epjc/s10052-019-6667-x}{Eur.\ Phys.\ J.\  \textbf{C79} (2019) 159}, \href{http://arxiv.org/abs/1812.06963}{{\normalfont\ttfamily arXiv:1812.06963}}\relax
\mciteBstWouldAddEndPuncttrue
\mciteSetBstMidEndSepPunct{\mcitedefaultmidpunct}
{\mcitedefaultendpunct}{\mcitedefaultseppunct}\relax
\EndOfBibitem
\bibitem{CKMfitter2015}
CKMfitter group, J.~Charles {\em et~al.}, \ifthenelse{\boolean{articletitles}}{\emph{{Current status of the standard model CKM fit and constraints on \hbox{$\Delta F=2$} new physics}}, }{}\href{https://doi.org/10.1103/PhysRevD.91.073007}{Phys.\ Rev.\  \textbf{D91} (2015) 073007}, \href{http://arxiv.org/abs/1501.05013}{{\normalfont\ttfamily arXiv:1501.05013}}, {updated results and plots available at \href{http://ckmfitter.in2p3.fr/}{{\texttt{http://ckmfitter.in2p3.fr/}}}}\relax
\mciteBstWouldAddEndPuncttrue
\mciteSetBstMidEndSepPunct{\mcitedefaultmidpunct}
{\mcitedefaultendpunct}{\mcitedefaultseppunct}\relax
\EndOfBibitem
\bibitem{HFLAV:2022pwe}
HFLAV collaboration, Y.~Amhis {\em et~al.}, \ifthenelse{\boolean{articletitles}}{\emph{{Averages of $b$-hadron, $c$-hadron, and $\tau$-lepton properties as of 2021}}, }{}\href{http://arxiv.org/abs/2206.07501}{{\normalfont\ttfamily arXiv:2206.07501}}\relax
\mciteBstWouldAddEndPuncttrue
\mciteSetBstMidEndSepPunct{\mcitedefaultmidpunct}
{\mcitedefaultendpunct}{\mcitedefaultseppunct}\relax
\EndOfBibitem
\bibitem{LHCb-PAPER-2020-036}
LHCb collaboration, R.~Aaij {\em et~al.}, \ifthenelse{\boolean{articletitles}}{\emph{{Measurement of \CP observables in $B^\pm \to D^{(*)} K^{\pm}$ and $B^\pm \to D^{(*)} \pi^{\pm} $ decays using two-body $D$ final states}}, }{}\href{https://doi.org/10.1007/JHEP04(2021)081}{JHEP \textbf{04} (2021) 081}, \href{http://arxiv.org/abs/2012.09903}{{\normalfont\ttfamily arXiv:2012.09903}}\relax
\mciteBstWouldAddEndPuncttrue
\mciteSetBstMidEndSepPunct{\mcitedefaultmidpunct}
{\mcitedefaultendpunct}{\mcitedefaultseppunct}\relax
\EndOfBibitem
\bibitem{Poluektov:2004mf}
Belle collaboration, A.~Poluektov {\em et~al.}, \ifthenelse{\boolean{articletitles}}{\emph{{Measurement of $\phi_3$ with Dalitz plot analysis of $B^{\pm} \rightarrow D^{(*)} K^{\pm}$ decay}}, }{}\href{https://doi.org/10.1103/PhysRevD.70.072003}{Phys.\ Rev.\  \textbf{D70} (2004) 072003}, \href{http://arxiv.org/abs/hep-ex/0406067}{{\normalfont\ttfamily arXiv:hep-ex/0406067}}\relax
\mciteBstWouldAddEndPuncttrue
\mciteSetBstMidEndSepPunct{\mcitedefaultmidpunct}
{\mcitedefaultendpunct}{\mcitedefaultseppunct}\relax
\EndOfBibitem
\bibitem{PhysRevLett.105.121801}
BaBar collaboration, P.~del Amo~Sanchez {\em et~al.}, \ifthenelse{\boolean{articletitles}}{\emph{Evidence for direct {CP} violation in the measurement of the {C}abbibo-{K}obayashi-{M}askawa angle $\ensuremath{\gamma}$ with ${B}^{\ensuremath{\mp}}\ensuremath{\rightarrow}{D}^{(*)}{K}^{(*)\ensuremath{\mp}}$ decays}, }{}\href{https://doi.org/10.1103/PhysRevLett.105.121801}{Phys.\ Rev.\ Lett.\  \textbf{105} (2010) 121801}, \href{http://arxiv.org/abs/1005.1096}{{\normalfont\ttfamily arXiv:1005.1096}}\relax
\mciteBstWouldAddEndPuncttrue
\mciteSetBstMidEndSepPunct{\mcitedefaultmidpunct}
{\mcitedefaultendpunct}{\mcitedefaultseppunct}\relax
\EndOfBibitem
\bibitem{Giri:2003ty}
A.~Giri, Y.~Grossman, A.~Soffer, and J.~Zupan, \ifthenelse{\boolean{articletitles}}{\emph{{Determining $\gamma$ using $B^{\pm} \rightarrow DK^{\pm}$ with multibody D decays}}, }{}\href{https://doi.org/10.1103/PhysRevD.68.054018}{Phys.\ Rev.\  \textbf{D68} (2003) 054018}, \href{http://arxiv.org/abs/hep-ph/0303187}{{\normalfont\ttfamily arXiv:hep-ph/0303187}}\relax
\mciteBstWouldAddEndPuncttrue
\mciteSetBstMidEndSepPunct{\mcitedefaultmidpunct}
{\mcitedefaultendpunct}{\mcitedefaultseppunct}\relax
\EndOfBibitem
\bibitem{Ablikim:2020lpk}
BESIII collaboration, M.~Ablikim {\em et~al.}, \ifthenelse{\boolean{articletitles}}{\emph{{Model-independent determination of the relative strong-phase difference between $D^0$ and $\bar{D}^0\rightarrow K^0_{S,L}\pi^+\pi^-$ and its impact on the measurement of the {CKM} angle $\gamma/\phi_3$}}, }{}\href{https://doi.org/10.1103/PhysRevD.101.112002}{Phys.\ Rev.\  \textbf{D101} (2020) 112002}, \href{http://arxiv.org/abs/2003.00091}{{\normalfont\ttfamily arXiv:2003.00091}}\relax
\mciteBstWouldAddEndPuncttrue
\mciteSetBstMidEndSepPunct{\mcitedefaultmidpunct}
{\mcitedefaultendpunct}{\mcitedefaultseppunct}\relax
\EndOfBibitem
\bibitem{Ablikim:2020cfp}
BESIII collaboration, M.~Ablikim {\em et~al.}, \ifthenelse{\boolean{articletitles}}{\emph{{Improved model-independent determination of the strong-phase difference between $D^{0}$ and $\bar{D}^0\to K^{0}_{\mathrm{S,L}}K^{+}K^{-}$ decays}}, }{}\href{https://doi.org/10.1103/PhysRevD.102.052008}{Phys.\ Rev.\  \textbf{D102} (2020) 052008}, \href{http://arxiv.org/abs/2007.07959}{{\normalfont\ttfamily arXiv:2007.07959}}\relax
\mciteBstWouldAddEndPuncttrue
\mciteSetBstMidEndSepPunct{\mcitedefaultmidpunct}
{\mcitedefaultendpunct}{\mcitedefaultseppunct}\relax
\EndOfBibitem
\bibitem{PhysRevD.82.112006}
CLEO collaboration, J.~Libby {\em et~al.}, \ifthenelse{\boolean{articletitles}}{\emph{Model-independent determination of the strong-phase difference between ${D}^{0}$ and $\bar{D}^0 \ensuremath{\rightarrow}{K}_{S,L}^{0}{h}^{+}{h}^{\ensuremath{-}}$ ($h=\ensuremath{\pi}$, $k$) and its impact on the measurement of the {CKM} angle $\ensuremath{\gamma}/{\ensuremath{\phi}}_{3}$}, }{}\href{https://doi.org/10.1103/PhysRevD.82.112006}{Phys.\ Rev.\  \textbf{D82} (2010) 112006}, \href{http://arxiv.org/abs/1010.2817}{{\normalfont\ttfamily arXiv:1010.2817}}\relax
\mciteBstWouldAddEndPuncttrue
\mciteSetBstMidEndSepPunct{\mcitedefaultmidpunct}
{\mcitedefaultendpunct}{\mcitedefaultseppunct}\relax
\EndOfBibitem
\bibitem{LHCb-PAPER-2020-019}
LHCb collaboration, R.~Aaij {\em et~al.}, \ifthenelse{\boolean{articletitles}}{\emph{{Measurement of the CKM angle $\gamma$ in \mbox{$B^{\pm} \to D K^{\pm}$ and $B^{\pm} \to D \pi^{\pm}$} decays with $D \to \KS h^+h^-$}}, }{}\href{https://doi.org/10.1007/JHEP02(2021)169}{JHEP \textbf{02} (2021) 0169}, \href{http://arxiv.org/abs/2010.08483}{{\normalfont\ttfamily arXiv:2010.08483}}\relax
\mciteBstWouldAddEndPuncttrue
\mciteSetBstMidEndSepPunct{\mcitedefaultmidpunct}
{\mcitedefaultendpunct}{\mcitedefaultseppunct}\relax
\EndOfBibitem
\bibitem{LHCb-PAPER-2018-017}
LHCb collaboration, R.~Aaij {\em et~al.}, \ifthenelse{\boolean{articletitles}}{\emph{{Measurement of the CKM angle $\gamma$ using \mbox{\decay{\Bpm}{D\Kpm}} with \mbox{\decay{D}{\KS \pip \pim,\ \KS \Kp\Km}} decays}}, }{}\href{https://doi.org/10.1007/JHEP08(2018)176}{JHEP \textbf{08} (2018) 176}, Erratum \href{https://doi.org/10.1007/JHEP10(2018)107}{ibid.\   \textbf{10} (2018) 107}, \href{http://arxiv.org/abs/1806.01202}{{\normalfont\ttfamily arXiv:1806.01202}}\relax
\mciteBstWouldAddEndPuncttrue
\mciteSetBstMidEndSepPunct{\mcitedefaultmidpunct}
{\mcitedefaultendpunct}{\mcitedefaultseppunct}\relax
\EndOfBibitem
\bibitem{LHCb-PAPER-2014-041}
LHCb collaboration, R.~Aaij {\em et~al.}, \ifthenelse{\boolean{articletitles}}{\emph{{Measurement of the CKM angle $\gamma$ using \mbox{\decay{\Bpm}{\D\Kpm}} with \mbox{\decay{\D}{\KS\pip\pim}}, $\KS\Kp\Km$ decays}}, }{}\href{https://doi.org/10.1007/JHEP10(2014)097}{JHEP \textbf{10} (2014) 097}, \href{http://arxiv.org/abs/1408.2748}{{\normalfont\ttfamily arXiv:1408.2748}}\relax
\mciteBstWouldAddEndPuncttrue
\mciteSetBstMidEndSepPunct{\mcitedefaultmidpunct}
{\mcitedefaultendpunct}{\mcitedefaultseppunct}\relax
\EndOfBibitem
\bibitem{Bondar:2004bi}
A.~Bondar and T.~Gershon, \ifthenelse{\boolean{articletitles}}{\emph{{On $\phi_{(3)}$ measurements using $B^{-} \rightarrow D^* K^-$ decays}}, }{}\href{https://doi.org/10.1103/PhysRevD.70.091503}{Phys.\ Rev.\  \textbf{D70} (2004) 091503}, \href{http://arxiv.org/abs/hep-ph/0409281}{{\normalfont\ttfamily arXiv:hep-ph/0409281}}\relax
\mciteBstWouldAddEndPuncttrue
\mciteSetBstMidEndSepPunct{\mcitedefaultmidpunct}
{\mcitedefaultendpunct}{\mcitedefaultseppunct}\relax
\EndOfBibitem
\bibitem{Bjorn:2019kov}
M.~Bj\o{}rn and S.~Malde, \ifthenelse{\boolean{articletitles}}{\emph{{CP violation and material interaction of neutral kaons in measurements of the CKM angle $\gamma$ using $B^\pm\to DK^\pm$ decays where $D\to K_\text{S}^0\pi^+\pi^-$}}, }{}\href{https://doi.org/10.1007/JHEP07(2019)106}{JHEP \textbf{07} (2019) 106}, \href{http://arxiv.org/abs/1904.01129}{{\normalfont\ttfamily arXiv:1904.01129}}\relax
\mciteBstWouldAddEndPuncttrue
\mciteSetBstMidEndSepPunct{\mcitedefaultmidpunct}
{\mcitedefaultendpunct}{\mcitedefaultseppunct}\relax
\EndOfBibitem
\bibitem{GarraTico:2018nng}
J.~Garra~Tic\'o, \ifthenelse{\boolean{articletitles}}{\emph{{A strategy for a simultaneous measurement of $CP$ violation parameters related to the $C\!K\!M$ angle $\gamma$ in multiple $B$ meson decay channels}}, }{}\href{http://arxiv.org/abs/1804.05597}{{\normalfont\ttfamily arXiv:1804.05597}}\relax
\mciteBstWouldAddEndPuncttrue
\mciteSetBstMidEndSepPunct{\mcitedefaultmidpunct}
{\mcitedefaultendpunct}{\mcitedefaultseppunct}\relax
\EndOfBibitem
\bibitem{PhysRevD.102.053003}
J.~Garra~Tic\'o {\em et~al.}, \ifthenelse{\boolean{articletitles}}{\emph{Study of the sensitivity to {CKM} angle $\ensuremath{\gamma}$ under simultaneous determination from multiple ${B}$ meson decay modes}, }{}\href{https://doi.org/10.1103/PhysRevD.102.053003}{Phys.\ Rev.\  \textbf{D102} (2020) 053003}, \href{http://arxiv.org/abs/1909.00600}{{\normalfont\ttfamily arXiv:1909.00600}}\relax
\mciteBstWouldAddEndPuncttrue
\mciteSetBstMidEndSepPunct{\mcitedefaultmidpunct}
{\mcitedefaultendpunct}{\mcitedefaultseppunct}\relax
\EndOfBibitem
\bibitem{LHCb-DP-2008-001}
LHCb collaboration, A.~A. Alves~Jr.\ {\em et~al.}, \ifthenelse{\boolean{articletitles}}{\emph{{The \lhcb detector at the LHC}}, }{}\href{https://doi.org/10.1088/1748-0221/3/08/S08005}{JINST \textbf{3} (2008) S08005}\relax
\mciteBstWouldAddEndPuncttrue
\mciteSetBstMidEndSepPunct{\mcitedefaultmidpunct}
{\mcitedefaultendpunct}{\mcitedefaultseppunct}\relax
\EndOfBibitem
\bibitem{LHCb-DP-2014-002}
LHCb collaboration, R.~Aaij {\em et~al.}, \ifthenelse{\boolean{articletitles}}{\emph{{LHCb detector performance}}, }{}\href{https://doi.org/10.1142/S0217751X15300227}{Int.\ J.\ Mod.\ Phys.\  \textbf{A30} (2015) 1530022}, \href{http://arxiv.org/abs/1412.6352}{{\normalfont\ttfamily arXiv:1412.6352}}\relax
\mciteBstWouldAddEndPuncttrue
\mciteSetBstMidEndSepPunct{\mcitedefaultmidpunct}
{\mcitedefaultendpunct}{\mcitedefaultseppunct}\relax
\EndOfBibitem
\bibitem{LHCb-DP-2014-001}
R.~Aaij {\em et~al.}, \ifthenelse{\boolean{articletitles}}{\emph{{Performance of the LHCb Vertex Locator}}, }{}\href{https://doi.org/10.1088/1748-0221/9/09/P09007}{JINST \textbf{9} (2014) P09007}, \href{http://arxiv.org/abs/1405.7808}{{\normalfont\ttfamily arXiv:1405.7808}}\relax
\mciteBstWouldAddEndPuncttrue
\mciteSetBstMidEndSepPunct{\mcitedefaultmidpunct}
{\mcitedefaultendpunct}{\mcitedefaultseppunct}\relax
\EndOfBibitem
\bibitem{LHCb-DP-2013-003}
R.~Arink {\em et~al.}, \ifthenelse{\boolean{articletitles}}{\emph{{Performance of the LHCb Outer Tracker}}, }{}\href{https://doi.org/10.1088/1748-0221/9/01/P01002}{JINST \textbf{9} (2014) P01002}, \href{http://arxiv.org/abs/1311.3893}{{\normalfont\ttfamily arXiv:1311.3893}}\relax
\mciteBstWouldAddEndPuncttrue
\mciteSetBstMidEndSepPunct{\mcitedefaultmidpunct}
{\mcitedefaultendpunct}{\mcitedefaultseppunct}\relax
\EndOfBibitem
\bibitem{LHCb-DP-2017-001}
P.~d'Argent {\em et~al.}, \ifthenelse{\boolean{articletitles}}{\emph{{Improved performance of the LHCb Outer Tracker in LHC Run 2}}, }{}\href{https://doi.org/10.1088/1748-0221/12/11/P11016}{JINST \textbf{12} (2017) P11016}, \href{http://arxiv.org/abs/1708.00819}{{\normalfont\ttfamily arXiv:1708.00819}}\relax
\mciteBstWouldAddEndPuncttrue
\mciteSetBstMidEndSepPunct{\mcitedefaultmidpunct}
{\mcitedefaultendpunct}{\mcitedefaultseppunct}\relax
\EndOfBibitem
\bibitem{LHCb-DP-2012-003}
M.~Adinolfi {\em et~al.}, \ifthenelse{\boolean{articletitles}}{\emph{{Performance of the \lhcb RICH detector at the LHC}}, }{}\href{https://doi.org/10.1140/epjc/s10052-013-2431-9}{Eur.\ Phys.\ J.\  \textbf{C73} (2013) 2431}, \href{http://arxiv.org/abs/1211.6759}{{\normalfont\ttfamily arXiv:1211.6759}}\relax
\mciteBstWouldAddEndPuncttrue
\mciteSetBstMidEndSepPunct{\mcitedefaultmidpunct}
{\mcitedefaultendpunct}{\mcitedefaultseppunct}\relax
\EndOfBibitem
\bibitem{LHCb-DP-2012-002}
A.~A. Alves~Jr.\ {\em et~al.}, \ifthenelse{\boolean{articletitles}}{\emph{{Performance of the LHCb muon system}}, }{}\href{https://doi.org/10.1088/1748-0221/8/02/P02022}{JINST \textbf{8} (2013) P02022}, \href{http://arxiv.org/abs/1211.1346}{{\normalfont\ttfamily arXiv:1211.1346}}\relax
\mciteBstWouldAddEndPuncttrue
\mciteSetBstMidEndSepPunct{\mcitedefaultmidpunct}
{\mcitedefaultendpunct}{\mcitedefaultseppunct}\relax
\EndOfBibitem
\bibitem{LHCb-DP-2012-004}
R.~Aaij {\em et~al.}, \ifthenelse{\boolean{articletitles}}{\emph{{The \lhcb trigger and its performance in 2011}}, }{}\href{https://doi.org/10.1088/1748-0221/8/04/P04022}{JINST \textbf{8} (2013) P04022}, \href{http://arxiv.org/abs/1211.3055}{{\normalfont\ttfamily arXiv:1211.3055}}\relax
\mciteBstWouldAddEndPuncttrue
\mciteSetBstMidEndSepPunct{\mcitedefaultmidpunct}
{\mcitedefaultendpunct}{\mcitedefaultseppunct}\relax
\EndOfBibitem
\bibitem{BBDT}
V.~V. Gligorov and M.~Williams, \ifthenelse{\boolean{articletitles}}{\emph{{Efficient, reliable and fast high-level triggering using a bonsai boosted decision tree}}, }{}\href{https://doi.org/10.1088/1748-0221/8/02/P02013}{JINST \textbf{8} (2013) P02013}, \href{http://arxiv.org/abs/1210.6861}{{\normalfont\ttfamily arXiv:1210.6861}}\relax
\mciteBstWouldAddEndPuncttrue
\mciteSetBstMidEndSepPunct{\mcitedefaultmidpunct}
{\mcitedefaultendpunct}{\mcitedefaultseppunct}\relax
\EndOfBibitem
\bibitem{Sjostrand:2007gs}
T.~Sj\"{o}strand, S.~Mrenna, and P.~Skands, \ifthenelse{\boolean{articletitles}}{\emph{{A brief introduction to PYTHIA 8.1}}, }{}\href{https://doi.org/10.1016/j.cpc.2008.01.036}{Comput.\ Phys.\ Commun.\  \textbf{178} (2008) 852}, \href{http://arxiv.org/abs/0710.3820}{{\normalfont\ttfamily arXiv:0710.3820}}\relax
\mciteBstWouldAddEndPuncttrue
\mciteSetBstMidEndSepPunct{\mcitedefaultmidpunct}
{\mcitedefaultendpunct}{\mcitedefaultseppunct}\relax
\EndOfBibitem
\bibitem{Sjostrand:2006za}
T.~Sj\"{o}strand, S.~Mrenna, and P.~Skands, \ifthenelse{\boolean{articletitles}}{\emph{{PYTHIA 6.4 physics and manual}}, }{}\href{https://doi.org/10.1088/1126-6708/2006/05/026}{JHEP \textbf{05} (2006) 026}, \href{http://arxiv.org/abs/hep-ph/0603175}{{\normalfont\ttfamily arXiv:hep-ph/0603175}}\relax
\mciteBstWouldAddEndPuncttrue
\mciteSetBstMidEndSepPunct{\mcitedefaultmidpunct}
{\mcitedefaultendpunct}{\mcitedefaultseppunct}\relax
\EndOfBibitem
\bibitem{LHCb-PROC-2010-056}
I.~Belyaev {\em et~al.}, \ifthenelse{\boolean{articletitles}}{\emph{{Handling of the generation of primary events in Gauss, the LHCb simulation framework}}, }{}\href{https://doi.org/10.1088/1742-6596/331/3/032047}{J.\ Phys.\ Conf.\ Ser.\  \textbf{331} (2011) 032047}\relax
\mciteBstWouldAddEndPuncttrue
\mciteSetBstMidEndSepPunct{\mcitedefaultmidpunct}
{\mcitedefaultendpunct}{\mcitedefaultseppunct}\relax
\EndOfBibitem
\bibitem{Lange:2001uf}
D.~J. Lange, \ifthenelse{\boolean{articletitles}}{\emph{{The EvtGen particle decay simulation package}}, }{}\href{https://doi.org/10.1016/S0168-9002(01)00089-4}{Nucl.\ Instrum.\ Meth.\  \textbf{A462} (2001) 152}\relax
\mciteBstWouldAddEndPuncttrue
\mciteSetBstMidEndSepPunct{\mcitedefaultmidpunct}
{\mcitedefaultendpunct}{\mcitedefaultseppunct}\relax
\EndOfBibitem
\bibitem{davidson2015photos}
N.~Davidson, T.~Przedzinski, and Z.~Was, \ifthenelse{\boolean{articletitles}}{\emph{{PHOTOS interface in C++: Technical and physics documentation}}, }{}\href{https://doi.org/https://doi.org/10.1016/j.cpc.2015.09.013}{Comp.\ Phys.\ Comm.\  \textbf{199} (2016) 86}, \href{http://arxiv.org/abs/1011.0937}{{\normalfont\ttfamily arXiv:1011.0937}}\relax
\mciteBstWouldAddEndPuncttrue
\mciteSetBstMidEndSepPunct{\mcitedefaultmidpunct}
{\mcitedefaultendpunct}{\mcitedefaultseppunct}\relax
\EndOfBibitem
\bibitem{Allison:2006ve}
Geant4 collaboration, J.~Allison {\em et~al.}, \ifthenelse{\boolean{articletitles}}{\emph{{Geant4 developments and applications}}, }{}\href{https://doi.org/10.1109/TNS.2006.869826}{IEEE Trans.\ Nucl.\ Sci.\  \textbf{53} (2006) 270}\relax
\mciteBstWouldAddEndPuncttrue
\mciteSetBstMidEndSepPunct{\mcitedefaultmidpunct}
{\mcitedefaultendpunct}{\mcitedefaultseppunct}\relax
\EndOfBibitem
\bibitem{Agostinelli:2002hh}
Geant4 collaboration, S.~Agostinelli {\em et~al.}, \ifthenelse{\boolean{articletitles}}{\emph{{Geant4: A simulation toolkit}}, }{}\href{https://doi.org/10.1016/S0168-9002(03)01368-8}{Nucl.\ Instrum.\ Meth.\  \textbf{A506} (2003) 250}\relax
\mciteBstWouldAddEndPuncttrue
\mciteSetBstMidEndSepPunct{\mcitedefaultmidpunct}
{\mcitedefaultendpunct}{\mcitedefaultseppunct}\relax
\EndOfBibitem
\bibitem{LHCb-PROC-2011-006}
M.~Clemencic {\em et~al.}, \ifthenelse{\boolean{articletitles}}{\emph{{The \lhcb simulation application, Gauss: Design, evolution and experience}}, }{}\href{https://doi.org/10.1088/1742-6596/331/3/032023}{J.\ Phys.\ Conf.\ Ser.\  \textbf{331} (2011) 032023}\relax
\mciteBstWouldAddEndPuncttrue
\mciteSetBstMidEndSepPunct{\mcitedefaultmidpunct}
{\mcitedefaultendpunct}{\mcitedefaultseppunct}\relax
\EndOfBibitem
\bibitem{Cowan:2016tnm}
G.~A. Cowan, D.~C. Craik, and M.~D. Needham, \ifthenelse{\boolean{articletitles}}{\emph{{RapidSim: an application for the fast simulation of heavy-quark hadron decays}}, }{}\href{https://doi.org/10.1016/j.cpc.2017.01.029}{Comput.\ Phys.\ Commun.\  \textbf{214} (2017) 239}, \href{http://arxiv.org/abs/1612.07489}{{\normalfont\ttfamily arXiv:1612.07489}}\relax
\mciteBstWouldAddEndPuncttrue
\mciteSetBstMidEndSepPunct{\mcitedefaultmidpunct}
{\mcitedefaultendpunct}{\mcitedefaultseppunct}\relax
\EndOfBibitem
\bibitem{Hocker:2007ht}
H.~Voss, A.~Hoecker, J.~Stelzer, and F.~Tegenfeldt, \ifthenelse{\boolean{articletitles}}{\emph{{TMVA - Toolkit for Multivariate Data Analysis with ROOT}}, }{}\href{https://doi.org/10.22323/1.050.0040}{PoS \textbf{ACAT} (2007) 040}\relax
\mciteBstWouldAddEndPuncttrue
\mciteSetBstMidEndSepPunct{\mcitedefaultmidpunct}
{\mcitedefaultendpunct}{\mcitedefaultseppunct}\relax
\EndOfBibitem
\bibitem{TMVA4}
A.~Hoecker {\em et~al.}, \ifthenelse{\boolean{articletitles}}{\emph{{TMVA 4 --- Toolkit for Multivariate Data Analysis with ROOT. Users Guide.}}, }{}\href{http://arxiv.org/abs/physics/0703039}{{\normalfont\ttfamily arXiv:physics/0703039}}\relax
\mciteBstWouldAddEndPuncttrue
\mciteSetBstMidEndSepPunct{\mcitedefaultmidpunct}
{\mcitedefaultendpunct}{\mcitedefaultseppunct}\relax
\EndOfBibitem
\bibitem{Breiman}
L.~Breiman, J.~H. Friedman, R.~A. Olshen, and C.~J. Stone, {\em Classification and regression trees}, Wadsworth international group, Belmont, California, USA, 1984\relax
\mciteBstWouldAddEndPuncttrue
\mciteSetBstMidEndSepPunct{\mcitedefaultmidpunct}
{\mcitedefaultendpunct}{\mcitedefaultseppunct}\relax
\EndOfBibitem
\bibitem{AdaBoost}
Y.~Freund and R.~E. Schapire, \ifthenelse{\boolean{articletitles}}{\emph{A decision-theoretic generalization of on-line learning and an application to boosting}, }{}\href{https://doi.org/10.1006/jcss.1997.1504}{J.\ Comput.\ Syst.\ Sci.\  \textbf{55} (1997) 119}\relax
\mciteBstWouldAddEndPuncttrue
\mciteSetBstMidEndSepPunct{\mcitedefaultmidpunct}
{\mcitedefaultendpunct}{\mcitedefaultseppunct}\relax
\EndOfBibitem
\bibitem{LHCb-DP-2020-001}
C.~Abellan~Beteta {\em et~al.}, \ifthenelse{\boolean{articletitles}}{\emph{{Calibration and performance of the LHCb calorimeters in Run 1 and 2 at the LHC}}, }{}\href{http://arxiv.org/abs/2008.11556}{{\normalfont\ttfamily arXiv:2008.11556}}, {submitted to JINST}\relax
\mciteBstWouldAddEndPuncttrue
\mciteSetBstMidEndSepPunct{\mcitedefaultmidpunct}
{\mcitedefaultendpunct}{\mcitedefaultseppunct}\relax
\EndOfBibitem
\bibitem{PDG2022}
Particle Data Group, R.~L. Workman {\em et~al.}, \ifthenelse{\boolean{articletitles}}{\emph{{\href{http://pdg.lbl.gov/}{Review of particle physics}}}, }{}\href{https://doi.org/10.1093/ptep/ptac097}{Prog.\ Theor.\ Exp.\ Phys.\  \textbf{2022} (2022) 083C01}\relax
\mciteBstWouldAddEndPuncttrue
\mciteSetBstMidEndSepPunct{\mcitedefaultmidpunct}
{\mcitedefaultendpunct}{\mcitedefaultseppunct}\relax
\EndOfBibitem
\bibitem{Skwarnicki:1986xj}
T.~Skwarnicki, {\em {A study of the radiative cascade transitions between the Upsilon-prime and Upsilon resonances}}, PhD thesis, Institute of Nuclear Physics, Krakow, 1986, {\href{http://inspirehep.net/record/230779/}{DESY-F31-86-02}}\relax
\mciteBstWouldAddEndPuncttrue
\mciteSetBstMidEndSepPunct{\mcitedefaultmidpunct}
{\mcitedefaultendpunct}{\mcitedefaultseppunct}\relax
\EndOfBibitem
\bibitem{LHCb-PAPER-2017-021}
LHCb collaboration, R.~Aaij {\em et~al.}, \ifthenelse{\boolean{articletitles}}{\emph{{Measurement of \CP observables in \mbox{\decay{\Bpm}{D^{(\ast)}\Kpm}} and \mbox{\decay{\Bpm}{D^{(\ast)}\pipm}} decays}}, }{}\href{https://doi.org/10.1016/j.physletb.2017.11.070}{Phys.\ Lett.\  \textbf{B777} (2018) 16}, \href{http://arxiv.org/abs/1708.06370}{{\normalfont\ttfamily arXiv:1708.06370}}\relax
\mciteBstWouldAddEndPuncttrue
\mciteSetBstMidEndSepPunct{\mcitedefaultmidpunct}
{\mcitedefaultendpunct}{\mcitedefaultseppunct}\relax
\EndOfBibitem
\bibitem{BaBar:2018cka}
BaBar, Belle collaborations, I.~Adachi {\em et~al.}, \ifthenelse{\boolean{articletitles}}{\emph{{Measurement of $\cos{2\beta}$ in $B^{0} \to D^{(*)} h^{0}$ with $D \to K_{S}^{0} \pi^{+} \pi^{-}$ decays by a combined time-dependent Dalitz plot analysis of BaBar and Belle data}}, }{}\href{https://doi.org/10.1103/PhysRevD.98.112012}{Phys.\ Rev.\  \textbf{D98} (2018) 112012}, \href{http://arxiv.org/abs/1804.06153}{{\normalfont\ttfamily arXiv:1804.06153}}\relax
\mciteBstWouldAddEndPuncttrue
\mciteSetBstMidEndSepPunct{\mcitedefaultmidpunct}
{\mcitedefaultendpunct}{\mcitedefaultseppunct}\relax
\EndOfBibitem
\bibitem{efron:1979}
B.~Efron, \ifthenelse{\boolean{articletitles}}{\emph{Bootstrap methods: Another look at the jackknife}, }{}\href{https://doi.org/10.1214/aos/1176344552}{Ann.\ Statist.\  \textbf{7} (1979) 1}\relax
\mciteBstWouldAddEndPuncttrue
\mciteSetBstMidEndSepPunct{\mcitedefaultmidpunct}
{\mcitedefaultendpunct}{\mcitedefaultseppunct}\relax
\EndOfBibitem
\bibitem{GammaCombo}
M.~Kenzie {\em et~al.}, \ifthenelse{\boolean{articletitles}}{\emph{{GammaCombo: A statistical analysis framework for combining measurements, fitting datasets and producing confidence intervals}}, }{}
\newblock doi:~\href{https://doi.org/10.5281/zenodo.3371421}{10.5281/zenodo.3371421}\relax
\mciteBstWouldAddEndPuncttrue
\mciteSetBstMidEndSepPunct{\mcitedefaultmidpunct}
{\mcitedefaultendpunct}{\mcitedefaultseppunct}\relax
\EndOfBibitem
\bibitem{matthew_kenzie_2021_5503651}
M.~Kenzie {\em et~al.}, \ifthenelse{\boolean{articletitles}}{\emph{gammacombo/gammacombo: {LHCb-PAPER}-2021-033}, }{} 2021.
\newblock doi:~\href{https://doi.org/10.5281/zenodo.5503651}{10.5281/zenodo.5503651}\relax
\mciteBstWouldAddEndPuncttrue
\mciteSetBstMidEndSepPunct{\mcitedefaultmidpunct}
{\mcitedefaultendpunct}{\mcitedefaultseppunct}\relax
\EndOfBibitem
\bibitem{LHCb-PAPER-2021-033}
LHCb collaboration, R.~Aaij {\em et~al.}, \ifthenelse{\boolean{articletitles}}{\emph{{Simultaneous determination of CKM~angle~$\gamma$ and charm mixing parameters}}, }{}\href{https://doi.org/10.1007/JHEP12(2021)141}{JHEP \textbf{12} (2021) 141}, \href{http://arxiv.org/abs/2110.02350}{{\normalfont\ttfamily arXiv:2110.02350}}\relax
\mciteBstWouldAddEndPuncttrue
\mciteSetBstMidEndSepPunct{\mcitedefaultmidpunct}
{\mcitedefaultendpunct}{\mcitedefaultseppunct}\relax
\EndOfBibitem
\end{mcitethebibliography}
